\documentclass[twocolumn]{autart}    % Enable this line and disable the 
\usepackage{graphicx}          % Include this line if your 
\usepackage{mathtools}
\usepackage{listings}  
\usepackage{float}
\usepackage{empheq}
\usepackage{caption}
\usepackage{subcaption}
\usepackage{multirow}
\usepackage{footnote}
\usepackage{algorithm}
\usepackage{algpseudocode}
\usepackage{subfiles}
\usepackage{amsfonts}
\usepackage{xcolor}
\usepackage{comment}

\newtheorem{lemma}{Lemma}
\newtheorem{theorem}{Theorem}
\newtheorem{proposition}{Proposition}
\newtheorem{corollary}{Corollary}

\theoremstyle{definition}
\newtheorem{definition}{Definition}
\newtheorem{example}{Example}
\theoremstyle{remark}
\newtheorem{remark}{Remark}

\begin{document}
	
	\begin{frontmatter}
		
		\title{Properties of Immersions for Systems with Multiple Limit Sets with Implications to Learning Koopman Embeddings\thanksref{footnoteinfo}} % Title, preferably not more 
		
		\thanks[footnoteinfo]{This research was supported in part by United States ONR grant  N00014-21-1-2431 and AFOSR grant FA9550-21-1-0289.}
		
		\author[First]{Zexiang Liu} 
		\author[First]{Necmiye Ozay} 
		\author[Second]{Eduardo D. Sontag}
		
		\address[First]{Department of Electrical Engineering and Computer Science, University of Michigan, Ann Arbor, MI 48109 USA \\(e-mail: \{zexiang, necmiye\}@umich.edu)}
		\address[Second]{
			Department of Electrical and Computer Engineering and Department
			of Bioengineering, Northeastern University, Boston, MA 02115 USA (e-mail: sontag@sontaglab.org)}

		                             % keyword list or with the 
		% help of the Automatica 
		% keyword wizard

		\begin{abstract}                          % Abstract of not more than 200 words.
			Linear immersions (such as Koopman eigenfunctions) of a nonlinear system have wide applications in prediction and control.	In this work, we study the properties of linear immersions for nonlinear systems with multiple omega-limit sets. While previous research has indicated the possibility of discontinuous one-to-one linear immersions for such systems, it has been unclear whether continuous one-to-one linear immersions are attainable. Under mild conditions, we prove that any continuous immersion to a class of systems including finite-dimensional linear systems collapses all the omega-limit sets, and thus cannot be one-to-one. Furthermore, we show that this property is also shared by approximate linear immersions learned from data as sample size increases and sampling interval decreases. Multiple examples are studied to illustrate our results. 
		\end{abstract}
		
	\end{frontmatter}
	
	\section{Introduction}
	Applied Koopman operator theory has drawn much attention in recent years due to its potential in the analysis, prediction, and control of nonlinear systems. The main idea behind this is fairly straightforward: As initially shown by \cite{koopman1931hamiltonian}, a nonlinear system can be equivalently represented by an infinite-dimensional linear system whose states consist of observables of the nonlinear system. If a finite-dimensional invariant subspace of this infinite-dimensional linear system is found, a finite-dimensional linear representation of the nonlinear system, called the \emph{Koopman representation}, can be extracted from its basis. This makes  prediction and control for nonlinear systems much easier since existing theoretical and algorithmic tools established for linear systems can now be applied to nonlinear systems via their finite-dimensional linear representations.  Compared with local linearization by Taylor expansion, the Koopman representation can capture global behaviors of the system \cite{mauroy2020koopman,brunton2021modern} and thus opens up exciting possibilities in various applications, such as model reduction and control of PDEs \cite{nathan2018applied,peitz2020feedback}, prediction of chaotic systems \cite{brunton2017chaos}, modeling and control of soft robots \cite{bruder2020data}, and model predictive control of nonlinear systems \cite{korda2018linear}.
	
	The idea behind Koopman representations and embeddings of nonlinear systems in linear (or bilinear, when there are controls) systems has been a recurring theme in the control literature, albeit under different names. Finite-dimensional embeddings correspond to finite-dimensional spaces of observables \cite{MR1025981}. The Koopman representation can be interpreted as the ``dual system'' used in linear theory (Kalman duality) and more generally as the foundation of the duality between observability of a nonlinear system and controllability of a (generally infinite-dimensional) system of observables, the \textit{adjoint system}. See for example the work in \cite{MR0479498,MR516861,icm94} on algebraic observability (strong reachability of the adjoint system, and surjective comorphisms into cosystems in the first reference) and a brief mention in Exercise 6.2.10 in the textbook \cite{mct}. A very closely related concept, but for infinite-dimensional linear systems, is ``topological observability'', which amounts to the exact reachability of a dual system \cite{yamamoto1981}.
	
	The primary challenge in applying Koopman operator theory to prediction and control lies in identifying a suitable Koopman representation. This involves finding a nonlinear transformation of the system states such that the transformed states evolve like a linear system. We call such a transformation \emph{a linear immersion}. Scalar-valued linear immersions are also known as Koopman eigenfunctions in the literature. As a trivial example, any constant function is a linear immersion for an arbitrary system, but this linear immersion is useless in practice since it does not include any information about the original nonlinear system. 
	Ideally, we want to find invertible linear immersions, ensuring that the trajectories of the original nonlinear system are fully characterized by its linear representation. 
	In instances where invertible linear immersions cannot be manually derived, especially for higher-dimensional systems, numerical approximation becomes necessary.  Various numerical methods have been developed to approximate linear immersions from data \cite{brunton2016discovering,tu2014dynamic,williams2015data}. A crucial guideline for achieving low approximation error in these methods is to carefully select a domain of interest where linear immersions are intended to be learned. In practice, for systems with multiple equilibria, a commonly mentioned insight is that a continuous one-to-one linear immersion across multiple isolated equilibria does not exist, supported by multiple analytical and numerical examples in the literature \cite{bakker2019learning,bakker2020learning,brunton2016koopman,mauroy2013spectral,page2019koopman,williams2015data}. Consequently, numerical methods are recommended to focus on learning local linear immersions within the domain of attraction of each equilibrium. However, recent work from \cite{arathoon2023koopman} challenges this insight. In particular, Arathoon and Kvalheim \cite{arathoon2023koopman} construct a smooth system with multiple isolated equilibria that admits a smooth one-to-one linear immersion. These positive and negative examples suggest that the non-existence of continuous one-to-one linear immersions is not solely determined by the presence of multiple isolated equilibria. To provide more accurate guidance on approximating linear immersions, it is imperative to reassess the aforementioned insight and identify the actual factors that determine the existence or non-existence of continuous one-to-one linear immersions. 
	
	To address these inquiries, in this work, we study the properties of continuous linear immersions for systems with multiple  limit sets, and their implications on algorithms that approximate linear immersions from data. In particular, our contributions include:
	\begin{itemize}
		\item We introduce a novel class of systems, termed \emph{systems with closed basins}, including both finite-dimensional linear systems and incrementally stable systems (Section \ref{sec:prelim}).
		\item For systems with multiple $\omega$-limit sets, we prove that any continuous immersions into a  system with closed basins, under mild conditions,  collapses all the $\omega$-limit sets into one and thus can not be one-to-one. We then demonstrate the applicability of our results with multiple examples from the literature (Section \ref{sec:main_thm}).
		\item For the same class of systems, we show that approximate linear immersions learned with data converge to functions that collapse all the $\omega$-limit sets, as sampling time decreases and sample size increases (Section \ref{sec:learning}).
		\item We show several extensions of the main theorem that can work with a broader class of systems (Section \ref{sec:exts}).
	\end{itemize}
	
	A preliminary version of this work was presented at the IFAC World Congress \cite{liu2023ifac},  focusing exclusively on one-to-one immersions. In this work, we extend the results in \cite{liu2023ifac} to encompass immersions that are not necessarily one-to-one in Section \ref{sec:main_thm}. Additionally, we introduce entirely new results in Sections \ref{sec:learning} and \ref{sec:exts}. 
	
	\textbf{Related work:} There is a rich literature on the classes of systems that can be immersed into linear systems, including systems confined in the domain of attraction of a stable equilibrium \cite{grune1999asymptotic,lan2013linearization} or a closed orbit \cite{lan2013linearization}, poly-flow systems \cite{van1994locally}, observable nonlinear systems \cite{levine1986nonlinear}, and even control-affine systems (for bilinear immersions) \cite{lo1975global}. However, the existence of linear immersions for systems with multiple isolated equilibria remains unresolved. Given that the presence of multiple equilibria is a key feature distinguishing nonlinear systems from linear ones, there has been significant discussion on whether such systems can be immersed into linear systems. It is initially observed in a numerical example from \cite{mauroy2013spectral} that the approximate linear immersion over a domain that contains two equilibria becomes singular at one of the equilibria.  Motivated by this observation, Williams et al. \cite{williams2015data} suggest that linear immersions should be approximated within the domain of attraction of each equilibrium to avoid singularities. This suggestion is supported by the work from  \cite{page2019koopman}, which studies a one-dimensional system with three equilibria where all the linear immersions can be derived manually. For this specific system, all the derived linear immersions become singular at one of the equilibria. Potentially motivated by these negative examples, Brunton et al. \cite{brunton2016koopman} claim that it is impossible to find one-to-one linear immersions for systems with multiple isolated equilibria. However, this claim is disproved by \cite{bakker2019learning}, which presents a one-dimensional system with three equilibria that admits a discontinuous one-to-one linear immersion. The paper \cite{bakker2019learning} further conjectures that a linear immersion may exist but become discontinuous at the boundaries of the basins of attraction.  This conjecture is again disproved by \cite{arathoon2023koopman}, which constructs a smooth system with two isolated equilibria that admits a smooth one-to-one linear immersion. Contrary to these varying claims, our work rigorously proves that continuous linear immersions cannot be one-to-one when the system has multiple isolated equilibria  and satisfies specific conditions. Our results confirm that the negative examples from the literature do not admit continuous one-to-one linear immersions, while the counter example in \cite{arathoon2023koopman} is the only one not meeting our extra conditions and thus allowing a continuous one-to-one linear immersion. Notably, \cite{kvalheim2023linearizability} provides necessary and sufficient conditions for the existence of one-to-one linear immersions for a class of nonlinear systems, which is different than the class of systems with multiple limit sets we consider in this paper.  While neither of these two classes is a subset of the other, for any system lying in the intersection of them, both our findings and those results from \cite{kvalheim2023linearizability} can infer the nonexistence of continuous one-to-one linear immersions. 
	
	\noindent \textbf{Notation}: We denote the closure of a set $X$ by $\text{cl}(X)$. The symbols $\mathbb{R}$, $\mathbb{R}_{\geq 0}$, and $\mathbb{R}_{>0}$ denote the real line, the set of non-negative real numbers, and the set of positive real numbers.  The symbols $\mathbb{Z}$ and $\mathbb{N}$ denote the set of integers and the set of non-negative integers. A function $\alpha:[0, a)\rightarrow \mathbb{R}_{\geq 0}$, for some $a>0$, belongs to class $\mathcal{K}$ if it is strictly increasing and $\alpha(0) = 0$.
	
	\section{Preliminaries} \label{sec:prelim} 
	\subsection{Problem Statement}
	We consider a continuous-time autonomous system {defined on a (second countable) manifold $\mathcal{M}$} :
	\begin{equation} \label{eq:sys} 
		\dot{x} = f(x),\  x\in \mathcal{M}.
	\end{equation}
	Given an initial state $\xi\in \mathcal{M}$, we denote the solution of the system in \eqref{eq:sys} by $\varphi:\mathbb{R}_{\geq 0}\times \mathcal{M} \rightarrow \mathcal{M} $ satisfying $\varphi(0,\xi) = \xi$ and for all $t\geq 0$,
	\begin{align}
		\frac{d\varphi(t,\xi)}{dt} = f(\varphi(t,\xi)).
	\end{align}
	Let $\mathcal{X}$ be a path-connected subset of the manifold $\mathcal{M}$ that represents the region in which we want to analyze the system behavior. We endow $\mathcal{X}$ with the subspace topology induced from $\mathcal{M}$. Throughout the paper, we will assume that $\varphi(t,\xi)$ is defined and contained in $\mathcal{X}$ for all $\xi\in \mathcal{X}$ and $t\geq 0$. Furthermore, $f$ is smooth enough to guarantee the uniqueness and continuous dependence on the initial states of the solution $\varphi(t,\xi)$ for all $\xi\in \mathcal{X}$ and all $t\geq 0$.
	
	\begin{remark}
		Every subspace of a second countable space, such as $\mathcal{X}$, is also second countable, which in turn implies that a subset of $\mathcal{X}$ is compact if and only if it is sequentially compact. We will use this last property.
	\end{remark}
	Given an initial state $\xi$ of {the} system in \eqref{eq:sys}, we denote the \emph{$\omega$-limit set}  of  $\xi$ in $\mathcal{X}$ by $\omega^{+}(\xi)$, that is the set of all $x\in \mathcal{X}$ satisfying that there exists a sequence $t_{n} \rightarrow \infty$ such that ${\lim_{n \to \infty} \varphi(t_{n},\xi) = x}$ \cite{hirsch2012differential}.
	
	Next, we introduce the definition of immersion, which generalizes the notion of Koopman eigenfunctions \cite{mauroy2020koopman}. 
	\begin{definition} \label{def:immersion} 
		A system $\dot{x} = f(x)$ on $\mathcal{X} \subseteq\mathcal{M}$ is \emph{immersed} in a system $\dot{z}=g(z)$ on a manifold $\mathcal{Z}$ if there is a mapping $F:\mathcal{X} \rightarrow \mathcal{Z}	$ (an immersion) such that, for all initial states $\xi\in \mathcal{X}$ and all time $t\geq 0$, 
		\begin{align}\label{eqn:immersion}
			F(\varphi(t,\xi))= \psi(t,F(\xi)),
		\end{align}	
		where $\psi(t,F(\xi))$ is the solution of $\dot{z}=g(z)$.  
		
		If the system $\dot{z} = g(z)$ above is linear, the mapping $F$ is called a \emph{linear immersion}.
	\end{definition}
	
	\begin{remark}\label{continuity}
		This paper investigates the properties of continuous immersions for systems with multiple $\omega$-limit sets. Throughout the remainder of this paper, unless otherwise specified, any immersion under consideration is assumed to be continuous. 
	\end{remark}	
	
	\begin{remark}\label{koopman operator}
		Linear immersions are tightly related to Koopman operator theory (\cite{brunton2016koopman}): A \emph{Koopman eigenfunction} $F$ is a (not necessarily continuous)  linear immersion that immerses $\dot{x} =f(x)$ in a one-dimensional system $\dot{z} = \lambda  z$ for some $ \lambda\in \mathbb{R}$. The span of the entries of any linear immersion $F$ is \emph{a Koopman invariant subspace}. 
		
	In the Koopman operator literature, the question of embeddings into finite-dimensional linear systems has been studied in the context of finite-dimensional spaces of observables invariant under the Koopman operator, as discussed in \cite{brunton2016koopman,sontag1995spaces}.  However, a finite-dimensional Koopman invariant subspace that fully characterizes the Koopman operator may not exist.  In our work, we establish the non-existence of immersions of an even more general form, not necessarily arising in this fashion.
	\end{remark}
	
	\begin{remark}\label{topological conjugacy} If an immersion $F$ is one-to-one, the inverse $F^{-1}: F(\mathcal{X}) \rightarrow \mathcal{X}$ exists. Thus  we can retrieve the solution $\varphi(t,\xi)$ of $\dot{x}=f(x)$ for any $\xi\in \mathcal{X}$ from the solution $\psi(t, F(\xi))$ of $\dot{z} = g(z)$ via the formula
		$\varphi(t,\xi) = F^{-1}(\psi(t, F(\xi)))$.
	\end{remark}
	
	\begin{remark}\label{immersion}
		The term ``immersion" is also widely used in the study of differentiable manifolds  (see for example \cite{lee2012smooth}), which is unrelated to the immersion of dynamical systems considered in this work.
	\end{remark}
	
	Given a nonlinear system $\dot{x}= f(x)$, we are most interested in finding a one-to-one linear immersion $F$, which fully encapsulates the behaviors of the nonlinear system into a linear system. However, in practice finding a one-to-one linear immersion can be very challenging, and sometimes such a linear immersion may not even exist. In particular, one might think that a one-to-one linear immersion may not exist when the $\omega$-limit sets of the nonlinear system are ``topologically" different from those of linear systems. For instance, nonlinear systems may have limit cycles but linear systems cannot.   However, the following example shows that it is possible to immerse a system with limit cycles into a  linear system.
	
	\begin{figure}[]
		\centering
		\includegraphics[width=0.3\textwidth]{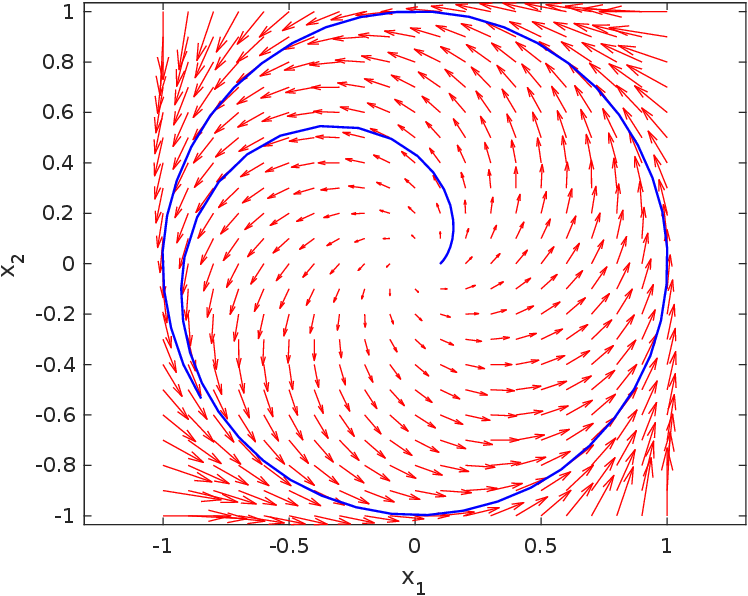}
		\caption{The vector field (red) of the system in \eqref{eq:sys_ex_3}.  The blue curve shows a trajectory of the system that starts from $(0.1,0)$ and converges to the unit circle.}
		\label{fig:ex_3}
	\end{figure}
	
	\begin{example}\label{ex_2D}
		Consider a two-dimensional system 
		\begin{align} \label{eq:sys_ex_3} 
			\begin{split}
				\dot{x}_{1}  &=  x_1-x_2 -x_1(x_1^{2}+x_2^{2}),\\
				\dot{x}_{2} &= x_1+x_2 - x_2(x_1^{2}+x_2^{2}).
			\end{split}
		\end{align}
		with state $x=(x_1,x_2)$. The system has an unstable equilibrium at the origin and one stable limit cycle on the unit circle, as shown by the phase portrait in Fig. \ref{fig:ex_3}. 
		Let $\mathcal{X} = \mathbb{R}^2 \backslash \{0\}$.  Intuitively, one may think a linear immersion does not exist for this system over $\mathcal{X}$ since linear systems cannot have a limit cycle. However, this system does admit a one-to-one linear immersion over $\mathcal{X}$. Let $F:\mathcal{X} \rightarrow \mathbb{R}^{3}$ be
		\begin{align}
			\label{eq:F_ex_3_2}
			F(x) = (x_1/\Vert x\Vert_2,\ x_2/\Vert x\Vert_2,\ \Vert x\Vert_2^{-2}-1).
		\end{align}
		For a solution $x(t)$ of the system in \eqref{eq:sys_ex_3}, it can be checked that $F(x(t))$ is a solution of the following linear system
		\begin{align}
			\begin{split}
				\dot{u} &= -v, \\
				\dot{v} &= u,\\
				\dot{w} &= -2w,
			\end{split}.
		\end{align}
		Thus, the one-to-one function $F$ in \eqref{eq:F_ex_3_2} is a linear immersion of the two-dimensional system.
	\end{example}
	
	\begin{figure}[]
		\centering
		\includegraphics[width=0.5\textwidth]{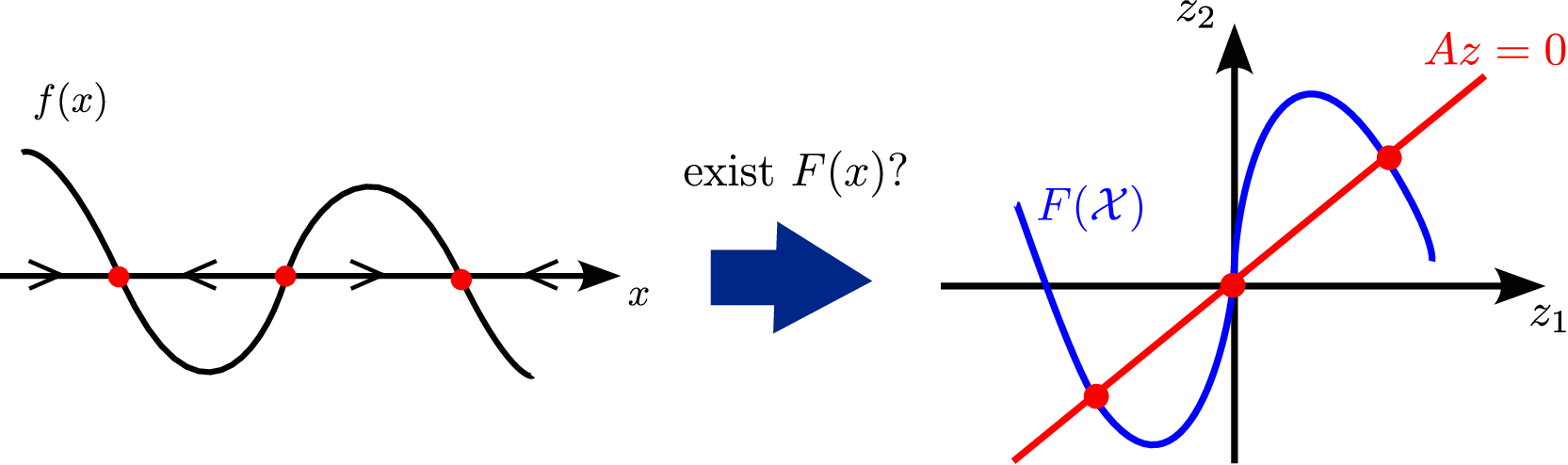}
		\caption{For a system $\dot{x} = f(x)$ with three equilibria, the graph of a one-to-one linear immersion  (blue) must intersect with the equilibrium set (red) of the immersing linear system at exactly three points.}
		\label{fig:koopman_snake}
	\end{figure}
	
	This paper focuses on another well-known topological difference between linear and nonlinear systems, namely that a nonlinear system can have multiple isolated $\omega$-limit sets, but a linear system cannot. We wonder how the properties of linear immersions, such as injectivity, are influenced by the presence of multiple $\omega$-limit sets. 
	Many existing works \cite{mauroy2013spectral,brunton2016koopman,page2019koopman} claim or observe that a continuous one-to-one linear immersion does not exist when the system possesses multiple isolated equilibria, a specific type of $\omega$-limit sets. However, a formal analysis of this phenomenon is missing in the literature. The following discussion explains why it is nontrivial to prove this claim: Suppose that an immersion $F$ that maps a system $\dot{x}=f(x)$ with multiple equilibria to a linear system $\dot{z} = Az$ exists. According to \eqref{eqn:immersion}, a one-to-one $F$ must map equilibria of $\dot{x} =f(x)$ to equilibria of the linear system $\dot{z} = A z$, and map non-equilibrium points to non-equilibrium points. However, recall that a linear system can only have one isolated equilibrium or a subspace of equilibria. If $\dot{z} = Az$ is the former, then $F$ maps all equilibria of $\dot{x}= f(x)$ to the unique equilibrium of $\dot{z}= Az$ and thus can not be one-to-one. Thus, the main challenge of proving or disproving this claim is to show if it is possible to have a one-to-one $F$ that maps $\dot{x}= f(x)$ to an immersing system $\dot{z}= Az$ with a subspace of equilibria. In this case, the only possibility is that the graph of $F$ intersects with the null space of $A$ at exactly $M$ points, with $M$ the number of equilibria of $\dot{x}=f(x)$, as demonstrated by Fig. \ref{fig:koopman_snake}. The following example shows that this is indeed possible if we allow the immersion to be discontinuous. 
	
	\begin{example}\label{ex_one_dim} Consider a one-dimensional system $\dot{x} = f(x)$ with $M$ isolated equilibria $\{x_{e,i}\}_{i=1}^M$, where $x_{e,1}<x_{e,2}< \cdots < x_{e,M}$. Assume that the solution $\varphi(t,\xi)$ is defined for any $\xi$ and $t\in \mathbb{R}$ (and thus has no finite-time blow-up both forward and backward in time). Let $\mathcal{X} = \mathbb{R}$.  This system is immersed in the following two-dimensional linear system 
		\begin{align} \label{eqn:sys_2d_immersed} 
			\begin{bmatrix}
				\dot{u}\\ \dot{v} 
			\end{bmatrix} = 
			\begin{bmatrix}
				0 & 1\\
				0 & 0\\
			\end{bmatrix} 
			\begin{bmatrix}
				u\\v
			\end{bmatrix},
		\end{align}
		by a one-to-one discontinuous function $F$
		\begin{align} \label{eqn:F_1d}
			&F(x) =\\
			&\begin{cases}
				(x, 0) & f(x)=0,\\
				(\varphi^{-1}(x, x_{e,1}-1) ,1) & x < x_{e,1},\\
				\left((i+1)\varphi^{-1}\left(x, \bar{x}_{e,i} \right) ,i+1\right) & x\in (x_{e,i}, x_{e,i+1}),\\
				((M+1)\varphi^{-1}(x, x_{e,M}+1) ,M+1) & x > x_{e,M},\\
			\end{cases}\nonumber
		\end{align}
		where $\bar{x}_{e,i} =(x_{e,i}+x_{e,i+1})/2$, and for any $x$ and $\xi \in (x_{e,i},x_{e,i+1})$, the inverse function $\varphi^{-1}(x, \xi)$ is the time instance $t$ such that $\varphi(t, \xi)=x$, which uniquely exists since $x$ and $\xi$ lie on the same trajectory. 
		Intuitively, the equilibria of $\dot{x}=f(x)$ cut the real line into $M+1$ intervals, and the function $F$ in \eqref{eqn:F_1d} maps trajectories within intervals to different horizontal lines in the lifted space, indexed by the state $v$. 
		
		\begin{figure}
			\centering
			\includegraphics[width=0.7\linewidth]{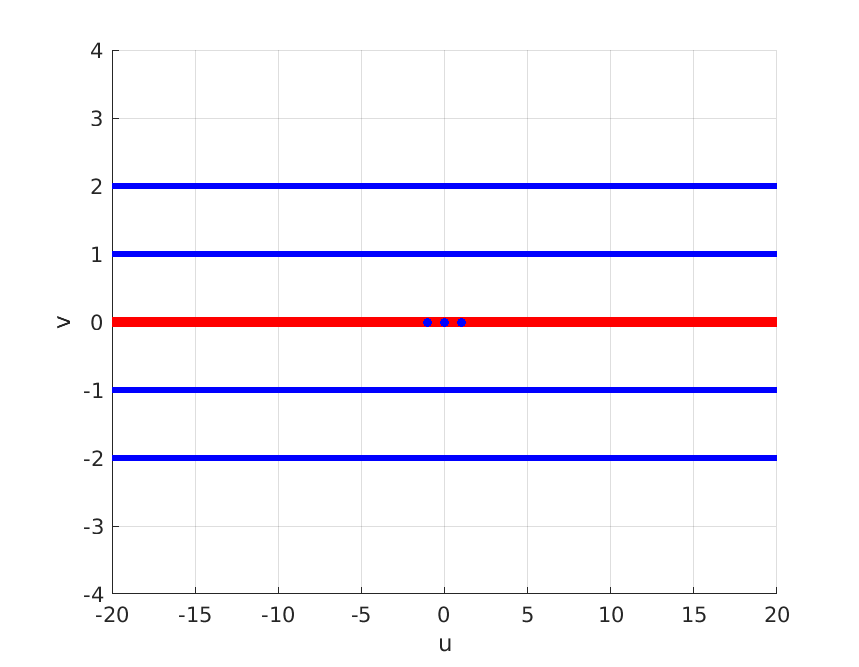}
			\caption[]{The image of the discontinuous immersion $F$ in \eqref{eqn:F_1d_3} is shown by the blue lines and dots. The set of equilibria of the immersed system in \eqref{eqn:sys_2d_immersed} is shown by the red line.}
			\label{fig:finalcurveimage}
		\end{figure}

		To see how this discontinuous immersion $F$ works in practice, we take a  
		concrete example of $\dot{x}=f(x)$ with $M=3$
		\begin{align} \label{eqn:sys_ex_1d} 
			\dot{x}= \frac{x ( 1- x^{2})}{1+x^{2}}.
		\end{align}
		The equilibria of this system include $\pm 1$ and $0$. According to \eqref{eqn:F_1d}, a one-to-one discontinuous linear immersion $F$ for this system is
		\begin{align} \label{eqn:F_1d_3} 
			F(x) = \begin{cases}
				(x,0) & f(x)=0,\\ 
				\left(\ln \bigg\vert \frac{3x}{-2+2x^{2}} \bigg\vert, 1\right) & x<-1,\\
				\left(2\ln \bigg\vert \frac{3x}{-2+2x^{2}} \bigg\vert, 2\right) & -1<x<0,\\
				\left(3\ln \bigg\vert \frac{3x}{-2+2x^{2}} \bigg\vert, 3\right) & 0<x<1,\\
				\left(4\ln \bigg\vert \frac{3x}{-2+2x^{2}} \bigg\vert, 4\right) & x>1.\\
			\end{cases}
		\end{align}
		The image of $F$ is shown in Fig. \ref{fig:finalcurveimage}. Let us examine the correctness of $F$ for $x_0\in (0,1)$. For any $x_0\in(0,1)$, the solution $\varphi(t,x_0)$ for $\dot{x}=f(x)$ in \eqref{eqn:sys_ex_1d} is
		\begin{align} \label{eqn:sol_1d} 
			\varphi(t,x_0) = \frac{x_0^{2}-1 + \sqrt{  \left( x_0^{2}-1 \right)^{2} + 4x_0^{2}e^{2t}}}{2x_0e^{t}}. 
		\end{align}
		Plugging the RHS of \eqref{eqn:sol_1d} into \eqref{eqn:F_1d_3}, we have 
		\begin{align}
			F(\varphi(t,x_0)) &= \left( 3 \ln \left( \frac{3x_0}{2-2x_0^{2}}  \right) +3t, 3 \right). \label{eqn:F_phi_t} 
		\end{align}
		According to \eqref{eqn:F_phi_t}, $F(\varphi(t,x_0))$ is equal to the solution of the linear system in \eqref{eqn:sys_2d_immersed} with respect to the initial state $F(x_0)$, showing the mapping $F$ in \eqref{eqn:F_1d_3} is a one-to-one linear immersion of \eqref{eqn:sys_ex_1d} over $(0,1)$.
	\end{example}
	
	Note that the graph of $F$ in \eqref{eqn:F_1d} intersects with the subspace $\mathbb{R}\times \{0\}$ of equilibria of \eqref{eqn:sys_2d_immersed} at precisely $M$ points, thanks to the discontinuity.  This phenomenon is also observed in continuous one-to-one linear immersions, exemplified in \cite{arathoon2023koopman}. This work aims to elucidate the relation between the properties of linear immersions, such as  injectivity and continuity, and the occurrence of multiple $\omega$-limit sets, as indicated by these examples and the literature.
	
	\noindent\fbox{%
		\parbox{0.45\textwidth}{%
			\textbf{Problem:} Identify the properties of linear immersions for systems with multiple $\omega$-limit sets.
	}}
	
	\subsection{Technical Definitions}
	
	\begin{definition}\label{def:precompact}
		Given an initial state $\xi\in \mathcal{X}$, the trajectory $\varphi(t, \xi)$ is called \emph{precompact} in $\mathcal{X}$ if the closure of the set $\{\varphi(t,\xi) \mid t\geq 0\} $  is compact  with respect to the subspace topology on $\mathcal{X}$.
	\end{definition}
	The following lemma states sufficient (and necessary) conditions for the nonemptiness of $ \omega^{+}(\xi)$. 
	\begin{lemma}\label{lem:precompact} 
		For any $\xi\in \mathcal{X}$, the $\omega$-limit set $\omega^{+}(\xi)$ is nonempty if the trajectory  $\varphi(t,\xi)$ is precompact {in $\mathcal{X}$}. If the system is linear {with $\mathcal{M} = \mathbb{R}^n$ and $\mathcal{X}$ closed}, then the converse is also true. 
	\end{lemma}
	\begin{pf}
		{The forward implication is well known. We recall the standard proof here.} Suppose $\varphi(t,\xi)$ is precompact {in $\mathcal{X}$}. Let $t_{n}$ be a sequence such that  $ t_{n} \rightarrow \infty$. 
		By Definition \ref{def:precompact}, there exists a subsequence $t_{n_{k}}$ such that $\varphi(t_{n_{k}},\xi)$ converges to a point $x$ in the closure of $\{\varphi(t,\xi) \mid t \in \mathbb{R}_{\geq 0}\}$. Thus, $\omega^{+}(\xi)$ contains $x$ and is nonempty.
		
		Now suppose that the system in \eqref{eq:sys} is linear, that is, $\dot{x} = Ax$ for some $A\in \mathbb{R}^{n\times  n}$. If a solution $\varphi(t,\xi)$ of the linear system is unbounded, it can be shown that for all $M > 0$, there exists $t_M \geq 0$ such that $  \Vert \varphi(t,\xi)\Vert_{2} \geq M$ for all $t\geq t_M$. Then,  $ \omega^{+}(\xi)$  is empty since for any sequence  $t_{n} \rightarrow \infty$, $ \Vert \varphi(t_{n},\xi)\Vert_{2} \rightarrow \infty$. 
		Thus, if $ \omega^{+}(\xi)$ is nonempty,  $\varphi(t,\xi)$ is bounded and thus is precompact in $\mathbb{R}^n$. Since $\mathcal{X}$ is closed, the closure of $\varphi(t,\xi)$ in $\mathbb{R}^n$ is contained in $\mathcal{X}$, which implies that $\varphi(t,\xi)$ is precompact in $\mathcal{X}$.\hfill $\Box$ 
	\end{pf}
	
	\begin{definition} \label{def:DoA}
		For any subset $\Omega$ of $\mathcal{X}$, we define its \emph{domain of attraction} by
		\begin{align}\label{eqn:D_plus}
			D^+(\Omega) = \{\xi \in \mathcal{X} \mid \omega^{+}(\xi)= \Omega\}. 
		\end{align}
	\end{definition}
	By \eqref{eqn:D_plus}, if $\Omega$ is not an $\omega$-limit set, $D^+(\Omega)$ is empty.
	
	Finally, we introduce a class of systems with a special property of the domain of attraction $D^+(\Omega)$. This class includes all linear systems.  Later we show that this special property is the main reason why a one-to-one linear immersion may not exist for a system with more than one $\omega$-limit set.

	\begin{definition} \label{def:closed_basins} 
		Let $\mathcal{W}^+$ be the set of all nonempty $\omega$-limit sets of $\dot{x}=f(x)$ in $\mathcal{X}$.
		A system of the form \eqref{eq:sys} has \emph{closed basins} if the domain of attraction $D^+(\Omega)$ is closed for all $\omega$-limit sets $\Omega\in \mathcal{W}^+$. 
	\end{definition}
	
	The following lemma provides sufficient conditions for systems to have closed basins, which are satisfied by all finite-dimensional linear systems.
	\begin{lemma}\label{lem:closed_basins} Any system $\dot{x}=f(x)$ defined over a subset $\mathcal{X}$ of a normed space has closed basins if for any $\omega$-limit set $\Omega$ in $\mathcal{W}^{+}$, the following two conditions are satisfied 
		\begin{enumerate}
			\item[(C1)] For any $\xi\in \mathcal{D}^{+}(\Omega)$, $\varphi(t,\xi)$ is precompact in $\mathcal{X}$.
			\item[(C2)] The system is incrementally stable in the closure of $\mathcal{D}^{+}(\Omega)$. That is, there exists a function $\alpha$ of class $\mathcal{K}$ such that, for any two initial states $\xi_1$ and $\xi_2$ in $\emph{cl}(\mathcal{D}^{+}(\Omega))$ and for all $t\geq 0$, $ \Vert \varphi(t,\xi_1) - \varphi(t,\xi_2)\Vert\leq \alpha ( \Vert \xi_1-\xi_2\Vert)$. 
		\end{enumerate}
	\end{lemma}
	\begin{pf}
		Let $\Omega$ be an arbitrary  $\omega$-limit set in $\mathcal{W}^{+}$. Let $x$ be a limit point of $\mathcal{D}^{+}(\Omega)$. That is, there exists a sequence $x_{k}\in \mathcal{D}^{+}(\Omega)$ such that $x_{k} \rightarrow x	$ as $k \rightarrow + \infty$.  
		
		We first show that $\omega^{+}(x)$ is nonempty and includes $\Omega$. 
		Pick any point $p\in \Omega$. For each $x_{k}$, since $\omega^{+}(x_{k})= \Omega$, there exists a sequence $t_{k} \rightarrow +\infty$ such that $ \Vert \varphi(t_{k},x_{k})-p\Vert \leq 1/k$. According to (C2), since $x\in \text{cl}(\mathcal{D}^{+}(\Omega))$,
		\begin{align}
			\Vert \varphi(t_{k},x)-p\Vert &\leq \Vert \varphi(t_{k},x_{k})-p\Vert + \Vert \varphi(t_{k},x) - \varphi(t_{k},x_{k})\Vert \nonumber\\
			&\leq 1/k + \alpha (\Vert x-x_{k}\Vert) \rightarrow 0 \text{ as }k \rightarrow + \infty. \nonumber		 
		\end{align}
		Therefore, $ \varphi(t_{k},x)$ converges to $p$ and thus $p\in \omega^{+}(x)$. Since $p$ is picked arbitrarily, $\Omega \subseteq \omega^{+}(x)$.
		
		Next, we want to show that $\Omega$ includes $\omega^{+}(x)$. We first prove the following claim: Given  $\xi_1$ and $\xi_2$ in $\text{cl}(D^{+}(\Omega))$ and any $p\in \omega^{+}(\xi_1)$, there is a $q\in\omega^{+}(\xi_2)$ such that $\Vert p-q\Vert \leq \alpha (\Vert \xi_1-\xi_2\Vert )$. 
		
		Let $t_{k} \rightarrow	+ \infty$ be a sequence such that $\varphi(t_{k},\xi_1) \rightarrow p$. By (C1), the sequence $\varphi(t_{k}, \xi_2)$ is contained in a compact subset of $\mathcal{X}$. Therefore, there exists a subsequence $t_{k}'$ of $t_{k}$ such that $\varphi(t_{k}',\xi_2) \rightarrow q$ for some $q$ in $\omega^{+}(\xi_2)$. By (C2), 
		\begin{align*}
			\Vert p-q\Vert  = \lim_{k \rightarrow + \infty	} \Vert \varphi(t_{k}',\xi_1) - \varphi(t_{k}', \xi_2)\Vert \leq \alpha( \Vert \xi_1-\xi_2\Vert ).
		\end{align*}
		
		Now we pick any point $p\in \omega^{+}(x)$. By the claim, there exists a sequence $q_{k}$ in $\omega^{+}(x_{k})$ such that $\Vert q_{k}-p\Vert \rightarrow 0$.  Since $\Omega$ is closed \cite{alligood2000chaos}, it follows that $p\in \Omega$. Since $p$ is arbitrary, $\omega^{+}(x)$ is a subset of $\Omega$.

		Since $\Omega \subseteq \omega^{+}(x)$ and $\omega^{+}(x) \subseteq \Omega$, we have $\omega^{+}(x) = \Omega$ and thus $x\in \mathcal{D}^{+}(\Omega)$.  Since $x$ is an arbitrary limit point of $\mathcal{D}^{+}(\Omega)$, $\mathcal{D}^{+}(\Omega)$ is closed. \hfill $\Box$
	\end{pf}
	\begin{remark}
		If $\mathcal{X}$ is a closed subset of a finite-dimensional normed space, the condition (C1) in Lemma \ref{lem:closed_basins}  can be replaced with the condition that for every $\Omega\in\mathcal{W}^+$, there exists one trajectory in $\mathcal{D}^+(\Omega)$ that is precompact in $\mathcal{X}$. We also note that we can define an abstract dynamical system over a metric space, and Lemma~\ref{lem:closed_basins} can be generalized accordingly.
	\end{remark}
	\begin{corollary} \label{cor:closed_basins_linear} 
		Every linear system $\dot{x}=Ax$ (with $\mathcal{X}=\mathbb{R}^{n}$) has closed basins. 
	\end{corollary}
	\begin{pf} 
		We want to prove the corollary by showing that any linear system satisfies the conditions (C1) and (C2) in Lemma \ref{lem:closed_basins}. The condition (C1) holds for any linear system according to Lemma \ref{lem:precompact}. 
		
		To show (C2), let $\Omega$ be an arbitrary $\omega$-limit set in $\mathcal{W}^+$. Denote the span of $D^+(\Omega)$ by $S$. By the superposition property of linear systems, since $D^+(\Omega)$ is forward invariant, $S$ is also forward invariant.  Thus, without loss of generality, we can restrict the state space of the system to $S$. According to (C1) and the superposition property of linear systems, all trajectories with initial states in the span $S$ of $D^+(\Omega)$ are precompact. That implies the system restricted to $S$ is stable in the sense of Lyapunov. Thus, there exists $M>0$ such that $\Vert \exp(At)x\Vert_{2} \leq M \Vert x\Vert_{2} $ for all $x\in S$ and $t\geq 0$. 
		
		Now we pick any two states $\xi_1$ and $\xi_2$ in $\text{cl}(\mathcal{D}^+(\Omega))$. Since $S$ is the span of $\mathcal{D}^+(\Omega)$,  $S$ contains $\xi_1$, $\xi_2$, and $\xi_1-\xi_2$. Therefore, for any $t\geq 0$, 
		\begin{align}
			\Vert \exp(At)\xi_1 - \exp(At)\xi_2\Vert_{2} &= \Vert \exp(At)(\xi_1 -\xi_2)\Vert_{2}\nonumber\\
			&\leq M \Vert \xi_1 -\xi_2\Vert_{2}.
		\end{align}
		Hence, any linear system satisfies the condition (C2).\hfill $\Box$
	\end{pf}
	
	\section{Main Theorem} \label{sec:main_thm} 
	The following theorem states our main results:
	\begin{theorem}\label{main_theorem}
		Suppose that:
		\begin{enumerate}
			\item[(T1)]
			$\dot x=f(x)$ on $\mathcal{X}$ can be immersed in a system with closed basins by a continuous mapping $F$;
			\item[(T2)]
			trajectories of $\dot x=f(x)$ on $\mathcal{X}$ are precompact in $\mathcal{X}$;
			\item[(T3)]
			the set ${\mathcal{W}^+}$ is countable.
		\end{enumerate}
		Then the set  $\{F(\Omega) \mid {\Omega \in \mathcal{W}^+}\} $ has exactly one element.
	\end{theorem}
	The proof of this theorem can be found in Appendix \ref{sec:proof}.
	This theorem essentially states that if there are a countable number of $\omega$-limit sets and the trajectories of the system are precompact, any continuous function $F$ that immerses the system into one with closed basins (in particular, any linear immersion) collapses all $\omega$-limit sets. A direct consequence of this result is as follows.
	
	\begin{corollary}\label{cor:one-to-one}
		Suppose that (T1), (T2), and (T3) hold and $F$ is one-to-one, then $\mathcal{W}^+$ has exactly one element.
	\end{corollary}
	Combining Corollaries \ref{cor:closed_basins_linear} and \ref{cor:one-to-one}, the following corollary states a necessary condition for the existence of one-to-one linear immersions (or in general one-to-one immersions to systems with closed basins).
	\begin{corollary}\label{cor:linear_system}
		If $\mathcal{X}$ contains more than one, but at most countably many, $\omega$-limit sets and all trajectories in $\mathcal{X}$ are precompact, then a one-to-one linear immersion does not exist for $\dot{x}=f(x)$ on $\mathcal{X}$.
	\end{corollary}
	
	For non-existence of linear immersions as in Corollary~\ref{cor:linear_system} both precompactness of trajectories and the existence of countable but more than one $\omega$-limit sets are not only sufficient conditions, they are indeed necessary in the following sense.  The paper
	\cite{arathoon2023koopman} provides an example of a two-dimensional system with two isolated equilibria with some trajectories that are neither precompact nor backward precompact (cf. Section~\ref{sec:direct}) that admits a linear immersion. Similarly, there are systems with uncountably many $\omega$-limit sets that admit linear immersions, simplest examples being diffeomorphisms of linear systems with a nontrivial subspace as their equilibria.
	
	Next, we demonstrate the application of our results through examples. We first show several examples where a one-to-one (linear) immersion is constructed manually when $\mathcal{X}$ {does not satisfy} the conditions in Corollary \ref{cor:linear_system}, but these immersions become discontinuous or ill-defined when we slightly modify $\mathcal{X}$ to violate one of these conditions. Sequentially, we offer examples from the literature where the existence of a linear immersion is unclear, but our results establish that continuous one-to-one linear immersions do not exist.
	
	\begin{example}\label{ex_1}
		Consider the one-dimensional system 
		\begin{align} \label{eq:sys_ex_1} 
			\dot{x} = x^{2}-1.
		\end{align}
		The $\omega$-limit sets of the system are  $\{-1\}$ and $\{1\}$.  Let $\mathcal{X}= (- \infty, 1)$, which only contains one $\omega$-limit set $\{-1\}$. It can be shown that $\dot{x}=x^{2}-1$ on $\mathcal{X}$ is immersed in $\dot{z} = -2z$ by the one-to-one mapping 
		\begin{align} \label{eq:F_ex_1} 
			F(x) = \frac{x+1}{x-1}. 
		\end{align}
		However, if we extend $\mathcal{X}$ by a point to $\mathcal{X}'=(- \infty, 1]$, the function $F$ in \eqref{eq:F_ex_1} is not an immersion anymore, since $F(1)$ is not defined. This observation is explained by Corollary \ref{cor:linear_system}: Since $\mathcal{X}'$ contains two $\omega$-limit sets $\{-1\} $ and $\{+1\} $, and all trajectories in $(- \infty,1]$ are precompact, there does not exist a one-to-one linear immersion for the system on $\mathcal{X}'$.
	\end{example}
	
	\begin{example}\label{ex_2}
		Consider the one-dimensional system:
		\begin{align} \label{eq:sys_ex_2} 
			\dot{x} = \sin(x).
		\end{align}
		Let $\mathcal{X}=[0, \pi]$. The $\omega$-limit sets of the system are $\{0\}$ and $\{\pi\}$.
		Define $y=\cos(x)$. Then, the derivative of $y$ satisfies 
		\begin{align} \label{eq:sys_ex_2_1} 
			\dot{y}  = - \sin(x)^{2} = \cos(x)^{2}-1 = y^{2}-1,
		\end{align}
		with $ \vert y \vert \leq 1$. That is, the system in \eqref{eq:sys_ex_2} on $\mathcal{X}$ is immersed in the system in \eqref{eq:sys_ex_1} on $\mathcal{Z} = [-1,1]$. In this example, $\mathcal{W}^+$ has two elements, and all trajectories of $x$ in $\mathbb{R}$ are precompact, but a one-to-one immersion exists. By Theorem \ref{main_theorem}, this is possible only if the system $\dot{y}=y^{2}-1$ does not have closed basins. Indeed, the domain of attraction  $D^+(\{-1\})$ of the system of $y$ on $\mathcal{Z}$ is $[-1,1)$, not a closed set.  
		
		Furthermore, by Example \ref{ex_1}, the system of $y$ on $[-1,1)$ can be immersed in $\dot{z} = -2z$ with the immersion in \eqref{eq:F_ex_1}. Thus, $\dot{x}= \sin(x)$ on $\mathcal{X}'= (0,\pi]$ is immersed in $\dot{z}=-2z$ with the one-to-one mapping
		\begin{align} \label{eq:F_ex_2} 
			F(x) = \frac{\cos(x)+1}{\cos(x)-1}.
		\end{align}
		If we extend $\mathcal{X}'$ to $\mathcal{X}=[0, \pi]$, the function $F(x)$ in \eqref{eq:F_ex_2} is not defined at $0$ and thus is not an immersion on the closed interval. This can be again explained by Corollary \ref{cor:linear_system} since all the trajectories of $x$ are precompact, and the interval $[0,\pi]$ contains two limit sets.
	\end{example}
	
	\begin{example}\label{ex_4}
		Consider the one-dimensional system
		\begin{align} \label{eq:sys_ex_4} 
			\dot{x} = x - x^{3}.
		\end{align}
		The $\omega$-limit sets of the system are $\{-1\} $, $\{0\} $ and $\{1\}$. Let $\mathcal{X} = \mathbb{R}\backslash \{0\} $. Define $y = x^{-2}-1$. Then, $\dot{y}$ satisfies
		\begin{align}
			\dot{y} = -2 x^{-3} (x-x^{3}) = -2y. 
		\end{align}
		Thus, the system in \eqref{eq:sys_ex_4} on $\mathcal{X}$ is immersed in $\dot{y}=-2y$ with the immersion $F(x)= x^{-2}-1$. Similar to the previous example, $\mathcal{X}$ contains two $\omega$-limit sets, but each of its path-connected components contains only one $\omega$-limit set and thus this observation is consistent with Corollary \ref{cor:linear_system}. 
	\end{example}
	
	\begin{example}\label{ex_3}
		Consider the two-dimensional system in Example \ref{ex_2D}. The $\omega$-limit sets of the system are the origin $\{0\}$ and the unit circle $\{x \mid \Vert x\Vert_{2}=1\} $.
		The linear immersion $F(x)$ in \eqref{eq:F_ex_3_2} is continuous in $\mathcal{X}$, but extending it to $\mathbb{R}^{2}$ makes $F(x)$ singular at the origin and thus no longer an immersion. This can be explained by Corollary \ref{cor:linear_system}: Since all trajectories of $x$ are precompact and the set $\mathbb{R}^{2}$ contains two $\omega$-limit sets, there does not exist a one-to-one linear immersion for the system of $x$ on $\mathbb{R}^{2}$.
	\end{example}
	
	\begin{example}\label{ex_one_dim_revisit}
		{Example \ref{ex_one_dim} shows that a discontinuous one-to-one linear immersion exists for the one-dimensional system in \eqref{eqn:sys_ex_1d} with three isolated equilibria. We question if this system possesses a continuous one-to-one linear immersion over $\mathbb{R}$.} This can be answered by Corollary \ref{cor:linear_system}: Since $\mathcal{X}$ contains three equilibria and all trajectories are precompact in $\mathcal{X}$, a continuous one-to-one linear immersion does not exist for this system.  
	\end{example}
	
\begin{figure}[]
		\centering
		\includegraphics[width=0.3\textwidth]{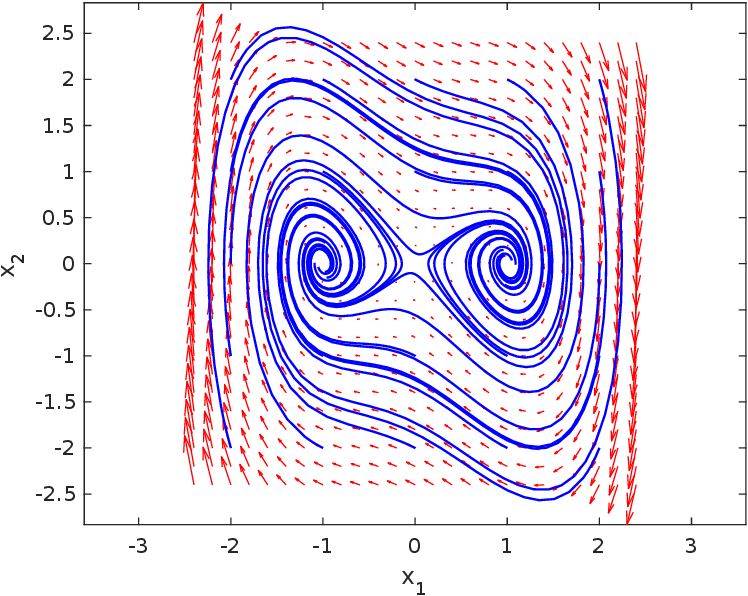}
		\caption{The vector field (red) and phase portrait (blue) of the unforced Duffing system in \eqref{eq:sys_duffing}. }
		\label{fig:ex_duffing}
	\end{figure}

	\begin{example}\label{ex_duffing}
		Consider the unforced Duffing system  in \cite{williams2015data}
		\begin{align} \label{eq:sys_duffing} 
			\begin{split}
				\dot{x}_{1} &= x_2\\
				\dot{x}_{2} &= -0.5 x_2 - x_1 (x_1^{2} -1).
			\end{split}
		\end{align}
		This system has two asymptotically stable equilibria $(\pm 1,0)$ and one unstable equilibrium $(0,0)$, as shown by Fig. \ref{fig:ex_duffing}.  Let $\mathcal{X}$ be the entire plane. The work in \cite{williams2015data} suggests that this system does not admit a one-to-one linear immersion over $\mathcal{X}$, which is confirmed by our results. According to Corollary \ref{cor:linear_system}, since the system has three $\omega$-limit sets and all of its trajectories are precompact in $\mathcal{X}$, there does not exist a one-to-one linear immersion over $\mathcal{X}$.
	\end{example}
	
\begin{figure}[]
		\centering
		\includegraphics[width=0.3\textwidth]{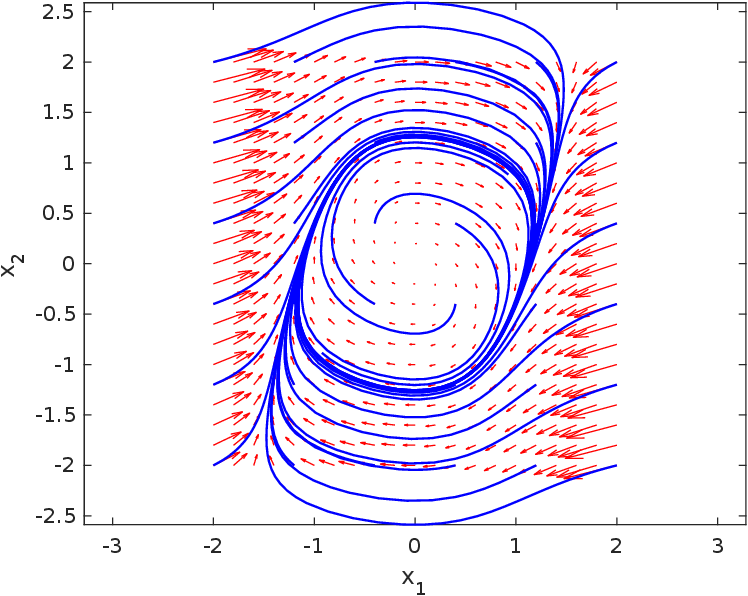}
		\caption{The vector field (red) and phase portrait (blue) of the Van der Pol equation in \eqref{eq:van_der_pol}.}
		\label{fig:ex_van_der_pol}
	\end{figure}

	\begin{example}\label{ex_van_der_pol}
		Consider the Van der Pol equation
		\begin{align}\label{eq:van_der_pol}
			\begin{split}
				\dot{x}_1 &= x_2 - x_1^3 + x_1\\
				\dot{x}_2 &= -x_1.
			\end{split}
		\end{align}
		This system has two $\omega$-limit sets, namely an unstable equilibrium at the origin and a stable limit cycle, as shown in Fig. \ref{fig:ex_van_der_pol}. Let $\mathcal{X}$ be the entire plane. Since all trajectories of the system are precompact in $\mathcal{X}$, by Corollary \ref{cor:linear_system}, there is no one-to-one linear immersion over $\mathcal{X}$.
	\end{example}
	
	\begin{example}\label{ex_lorenz}
		Consider the Lorenz system
		\begin{align}
			\begin{split}
				\dot{x} &=  \sigma(y-x)\\
				\dot{y} &= rx - y - xz\\
				\dot{z} &= xy - bz,
			\end{split}
		\end{align}
		with $ \sigma = 10$, $b = 8/3$, and $r = 28$. According to \cite{hirsch2012differential}, there exists an invariant ellipsoid $\mathcal{E}$ centered at $(0,0,2r)$ that contains all the $\omega$-limit sets of the system, which include three equilibria and the Lorenz attractor.  Let $\mathcal{X} = \mathcal{E}$.  Since $\mathcal{X}$ is invariant and compact, all trajectories are precompact in $\mathcal{X}$.  Thus, by Corollary \ref{cor:linear_system}, there does not exist a one-to-one linear immersion for the Lorenz system on $\mathcal{X}$.
	\end{example}
	
	\section{Implication for Learning Linear Immersions from Data} \label{sec:learning}
	In this section, we discuss the implications of our result for learning linear immersions of $\dot{x} =f(x)$ from data. 
	Throughout this section, we make the following assumption.
	\begin{assum} \label{asp:T2-4} 
		The system $\dot{x}=f(x)$ satisfies conditions (T2) and (T3) of Theorem \ref{main_theorem} on a path-connected forward-invariant subset $\mathcal{X}$ of $\mathcal{M}$. 
	\end{assum}
	
	Generally, given a fixed sampling time $\tau$, a set of $N$ pairs  $\{(x_{l},x_l^+)\}_{l=1}^{N}$ where $x_l^+ = \varphi(\tau,	x_l)$ for all $l$, and an integer $m>0$, the task of learning linear immersions $F: \mathcal{X} \rightarrow \mathbb{R}^m$ involves finding the following set
	\begin{align} 
		\mathcal{F}^{*}(\tau, N)=\{F^*\in \mathcal{C}(\mathcal{X}, \mathbb{R}^m) \mid \exists A^* \in \mathbb{R}^{m\times m}, \nonumber\\F^*(x_{l}^+) = e^{A^*\tau} F^*(x_{l}), \forall l=1, \cdots , N\},\label{eqn:learning} 
	\end{align}
	where $\mathcal{C}(\mathcal{X}, \mathbb{R}^m)$ is the space of continuous functions from $\mathcal{X}$ to $\mathbb{R}^m$.
	The state pairs $\{(x_{l}, x_{l}^{+})\}_{l=1}^{N}$ can be extracted from a single trajectory or multiple trajectories of the system. Essentially, $\mathcal{F}^{*}(\tau, N)$ is the set of continuous functions that satisfy the condition in \eqref{eqn:immersion} for linear immersions at finitely many points $\{x_{l}\}_{l=1}^{N}$ and a fixed time step $\tau$.
	
	Alternatively, the learning problem can be stated as:
	\begin{align} 
		(F^{*}, A^{*}) \in \underset{\substack{F\in \mathcal{C}( \mathcal{X}, \mathbb{R}^m),\\ A\in \mathbb{R}^{m\times m}}}{\arg\min} &~ \sum_{l=1}^{N} \Vert F(x_{l}') - e^{A\tau} F(x_{l})  \Vert \nonumber\\
		\text{s.t. } & x_{l}' = \varphi(\tau, x_{l}), \forall l\in\{1,\ldots,N\}. \label{eqn:learning_opt} 
	\end{align}
	The set $\mathcal{F}^{*}(\tau,N)$ in \eqref{eqn:learning} corresponds to the solutions $F^{*}$ of this problem that give zero objective value, i.e., those interpolating the data perfectly.

	Theorem \ref{main_theorem} shows that under Assumption \ref{asp:T2-4}, any linear immersion $F$ satisfies that for all pairs $\Omega_{i}$, $\Omega_{j}$ in $\mathcal{W}^+$,
	\begin{align}\label{eqn:F_image}
		F(\Omega_{i}) = F(\Omega_{j}).
	\end{align}
	A critical question here is if any learned linear immersion in $\mathcal{F}^{*}(\tau, N)$ would also share this property in \eqref{eqn:F_image}. As a sanity check, note that any constant function $F^*(\cdot)=C$ for some $C\in \mathcal{Z}$  belongs to $\mathcal{F}^*(\tau, N)$ (with the corresponding $A^*=0$) and does map every $\omega$-limit set to the same set. 
	However, this may not hold for every learned linear immersion in $\mathcal{F}^{*}(\tau, N)$ since the learned linear immersion only satisfies \eqref{eqn:immersion} at finitely many points (finitely many constraints), while a true linear immersion needs to satisfy \eqref{eqn:immersion} everywhere in $\mathcal{X}$ at all times (uncountably many constraints). {However, in Theorem \ref{theo:learning_asymp_2}, we show that any function that does not collapse all $\omega$-limit sets} would be excluded from $\mathcal{F}^{*}(\tau,N)$ for small enough $\tau$ and large enough $N$. The following proposition provides a crucial step for this result.
	
	\begin{proposition}\label{theorem_learning_asypm_2} 
		Let $\{x_l\}_{l=1}^\infty$ be a dense subset of $\mathcal{X}$. Let $F$ be a continuous mapping from $\mathcal{X}$ to $\mathbb{R}^m$.
		If for all $t> 0$, there exists  a sampling time $\tau\in (0,t]$ such that $F\in \mathcal{F}^*(\tau, N)$ for all $N > 1$, then $F$ is a linear immersion of the system.
	\end{proposition}
	
	\begin{pf}	
		Suppose that a continuous function $F$ satisfies the conditions in Theorem \ref{theorem_learning_asypm_2}. Then, for all $j\in \mathbb{N}$, there exists a sampling time $\tau_j\in (0,2^{-j}]$ such that $F\in \mathcal{F}^*(\tau_j, N)$ for all $N\geq 1$.  Clearly, the positive sequence $\tau_j$ converges to zero as $j$ goes to infinity. Also, for each $j$, $F\in \lim_{N\rightarrow \infty} \mathcal{F}^*(\tau_j, N) =: \mathcal{F}^*(\tau_j)$. Note that the limit $\mathcal{F}^*(\tau_j)$ is well defined since the set sequence $\mathcal{F}^*(\tau_j, N)$ monotonically shrinks as $N$ increases. 
		
		Next, for a fixed $j$, we want to prove that there exists a matrix $A$ such that for all $l\geq 1$, 
		\begin{align} \label{eqn:F_data} 
			F(\varphi(\tau_{j},x_{l})) = e^{A \tau_{j}} F(x_{l}).
		\end{align}
		We denote the set of matrices  $A_d$ such that $F(x_{l}^{+})=A_d F(x_{l})$ for all $l$ from $1$ to $N$ by $\mathcal{A}_d^{*}(N)$.   By definition, $\mathcal{A}_d^{*}(N)$ is an affine subspace in $\mathbb{R}^{m\times m}$, and monotonically shrinks with $N$. Since $\mathcal{A}_d^{*}(N)$ is an affine subspace for all $N \geq 1$, each time $\mathcal{A}_d^{*}(N)$ shrinks, its dimension decreases. Thus, the sequence of sets $\mathcal{A}_d^{*}(N)$ must converge at a finite $N^*$ since the dimension of $\mathcal{A}_d^{*}(N)$ can decrease at most finitely many times. 
		Since $F\in \mathcal{F}^{*}(\tau_{j}, N^{*})$, there exists at least one $A\in \mathbb{R}^{m\times m}$ 
		such that $e^{A\tau_j} \in \mathcal{A}_d^{*}(N^*)$. This matrix $A$ satisfies  \eqref{eqn:F_data} for all $ l\geq 1$. 
		
		Then, we pick an arbitrary $x\in \mathcal{X}$. Since  $\{x_{l}\}_{l=1}^{ \infty} $ is dense in $\mathcal{X}$, there exists a subsequence of $x_{l}$, denoted by $\overline{x}_{l}$, that converges to $x$. Since $F(\cdot)$ and $\varphi(\tau_{j},\cdot)$ are continuous,
		\begin{align}
			\label{eqn:F_all_x_2} 
			\begin{split}
				F(\varphi(\tau_{j},x)) &=  \lim_{l \to \infty} F(\varphi(\tau_{j},\overline{x}_{l}))\\
				&= \lim_{l \to \infty} e^{A\tau_{j}}F(\overline{x}_{l})\\
				&= e^{A\tau_{j}}F(x).
			\end{split}
		\end{align}
		We pick an arbitrary $t>0$. Since $\tau_j$ is positive and converging to zero (recall that $\tau_j\in (0, 2^{-j}]$), there exists $k_j\in \mathbb{N}$ such that the sequence $\bar{t}_j:=\sum_{i=1}^{j} k_j\cdot \tau_j$ converges to $t$ as $j$ goes to infinity. Note that for all $j$,  \eqref{eqn:F_all_x_2} implies that
		\begin{align}
			F(\varphi(\bar{t}_j,x)) = e^{A\bar{t}_j} F(x). \label{eqn:F_immersion} 
		\end{align}
		By the continuity of $F(\cdot)$ and $\varphi(\cdot, x)$, we have
		\begin{align}
			F(\varphi(t,x)) &= F(\varphi(\lim_{j\rightarrow \infty}\bar{t}_j,x))\nonumber \\
			&=\lim_{j\rightarrow \infty} F(\varphi(\bar{t}_j,x))\nonumber\\
			&=\lim_{j\rightarrow \infty} e^{A\bar{t}_j} F(x) = e^{A t} F(x). \label{eqn:F_immersion_2} 
		\end{align}
		Since $x$ and $t$ are arbitrary, \eqref{eqn:F_immersion_2} implies that $F$ is a linear immersion. 
		\hfill $\Box$
	\end{pf}
	
	Now, we state the main result of this section.
	
	\begin{theorem} \label{theo:learning_asymp_2} 
		Let $\{x_l\}_{l=1}^\infty$ be a dense subset of $\mathcal{X}$. Let $F$ be a continuous function such that $F(\Omega_1)\not=F(\Omega_2)$ for some $\Omega_1$ and $\Omega_2\in \mathcal{W}^{+}$. Then, there exists $t^* > 0$ such that for each sampling time $\tau\in (0,t^*]$, there exists an $N > 0$ such that $F\not\in \mathcal{F}^*(\tau, N)$.
	\end{theorem}
	\begin{pf}
		Let  $F$ be a continuous function that does not map every $\Omega\in \mathcal{W}^{+}$ to the same subset of $\mathbb{R}^{m}$.
		According to Theorem \ref{main_theorem}, since  (T2) and (T3) hold,  $F$ is not a linear immersion of the system. By the contrapositive of Proposition \ref{theorem_learning_asypm_2}, for any $F$ not a linear immersion, there exists $t>0$ such that for all $\tau\in (0,t]$ and for some large enough $N$, $F$ is not in $\mathcal{F}^{*}(\tau, N)$. \hfill $\Box$
	\end{pf}
	
	Theorem \ref{theo:learning_asymp_2} reveals that any immersion candidate $F$ that can distinguish at least two $\omega$-limit sets in $\mathcal{X}$  would always be ruled out from $\mathcal{F}(\tau, N)$ by a small enough sampling time $\tau$ and a large enough sample size $N$. This is particularly the case for common algorithms that learn Koopman embeddings using a continuous parameterization, such as polynomials \cite{williams2015data} and deep neural networks \cite{yeung2019learning}. Hence, these algorithms will suffer from the issues identified in this section as long as they try to minimize the cost in \eqref{eqn:learning_opt} and achieve zero fitting error.
	
	\begin{remark}\label{rm:conv}
		Using similar arguments in the proof of Theorem \ref{theorem_learning_asypm_2}, one can also show that
		for any positive time sequence $\tau_{j}$ that converges to zero, we have
		\begin{align}
			\limsup_{j\rightarrow \infty} \lim_{N\rightarrow \infty}\mathcal{F}^*(\tau_j, N) \subseteq \mathcal{C}(\mathcal{W}^+),
		\end{align}
		where $ \mathcal{C}(\mathcal{W}^+) $ is the set of functions $F\in \mathcal{C}(\mathcal{X}, \mathbb{R}^m) $ such that $ F(\Omega_1)=F(\Omega_2) $ for any $ \Omega_1 $ and $ \Omega_2\in \mathcal{W}^{+}$.
	\end{remark}
	
	\begin{remark} \label{rm:dense} 
		The condition of sampled states $\{x_{l}\}_{l=1}^{ \infty}$ being dense in $\mathcal{X}$ in Theorem \ref{theorem_learning_asypm_2} indicates that the data collection process is conducted in a way such that the domain $\mathcal{X}$ of interest is thoroughly covered by the sampled states in the limit. This condition is relatively straightforward to meet. For instance, consider a Borel probability measure $\mu$ over $\mathcal{X}$ such that any open subset of $\mathcal{X}$ is not measure zero.
		By repeatedly drawing random initial states according to $\mu$, simulating trajectories for a finite time horizon, and extracting state pairs $(x_{l},x_{l}^{+})$ from these trajectories, the resulting samples $\{x_{l}\}_{l=1}^{ \infty}$ is dense in $\mathcal{X}$ almost surely.
	\end{remark}
	
	\section{Extensions of the Main Theorem} \label{sec:exts}
	Due to the condition (T2), Theorem \ref{main_theorem} cannot be directly applied for systems with diverging trajectories in $\mathcal{X}$. In this section, we show several extensions to Theorem \ref{main_theorem} that accommodate systems with diverging trajectories.
	
	\subsection{Indirect Extensions of Theorem \ref{main_theorem}}
	The following two propositions, in conjunction with Theorem \ref{main_theorem}, show the non-existence of one-to-one linear immersions even when a diverging trajectory is present. We refer to these propositions as ``indirect extensions" to Theorem 1 because they must be utilized in conjunction with it.
	
	\begin{proposition} \label{prop:extend_X}
		If $F:\mathcal{X} \rightarrow \mathcal{Z}$ is an immersion of the system $\dot{x} = f(x)$ over $\mathcal{X}$, then $F$ is an immersion over any forward invariant subset of $\mathcal{X}$. 
	\end{proposition}
	The proof of Proposition \ref{prop:extend_X} is straightforward and omitted for brevity. By Proposition \ref{prop:extend_X}, if we show there is no one-to-one linear immersion over a forward invariant subdomain of $\mathcal{X}$, that implies a one-to-one linear immersion does not exist over $\mathcal{X}$. 
	
	\begin{proposition} \label{prop:time_reverse}
		Suppose that $\mathcal{X}' \subseteq\mathcal{X}$ is forward invariant for the time-reversed system $\dot{x} = -f(x)$. If $F$ immerses the system $\dot{x} = f(x)$ over $\mathcal{X}$ into $\dot{z} =g(z)$ over $\mathcal{Z}$, then the same $F$ also immerses the time-reversed system $ \dot{x} = -f(x)$ over $\mathcal{X}'$ into $\dot{z} = -g(z)$ over $F(\mathcal{X}')$.
	\end{proposition}
	\begin{pf}
		We want to show that the time-reversed system $\dot{x} =-f(x)$ over $\mathcal{X}'$ is immersed into $\dot{z} = -g(z)$ by $F$, which is equivalent to show that $F(\varphi(-t, \xi)) = \psi(-t, F(\xi))$ for all $\xi\in \mathcal{X}'$ and for all $t \geq 0$.
		
		Pick an arbitrary $\xi\in \mathcal{X}'$. Since $\mathcal{X}'$ is forward invariant for $\dot{x} = -f(x)$, $\varphi(-t,\xi)$ is well-defined for all $t\geq 0$. We denote $\mathcal{X}_{\xi} := \{\varphi(-t,\xi) \mid  t\in \mathbb{R}_{\geq 0}\} \subseteq \mathcal{X}'$.  We pick any $x\in \mathcal{X}_{\xi}$. There exists $\tau(x)\leq 0$ such that $x = \varphi(\tau(x), \xi)$. Since $F$ is an immersion over $\mathcal{X}$ and $x\in \mathcal{X}$, we have for all $t  \geq 0$,
		\begin{align}\label{eqn:F_phi}
			F(\varphi(t,x)) = \psi(t, F(x)).
		\end{align}
		Thus, $F(\xi) = F(\varphi(-\tau(x),x)) = \psi(-\tau(x), F(x))$. Since $\varphi(t,x) = \varphi(t+\tau(x), \xi)$, by \eqref{eqn:F_phi},
		\begin{align} \label{eqn:F_psi} 
			\begin{split}
				F(\varphi(t+\tau(x),\xi)) &= \psi(t, F(x)) \\
				&= \psi(t+\tau(x), \psi(-\tau(x), F(x))) \\
				&= \psi(t+\tau(x), F(\xi)).
			\end{split}
		\end{align}
		Let $t=0$. Then \eqref{eqn:F_psi} implies   
		$ F(\varphi(\tau(x),\xi)) = \psi(\tau(x), F(\xi))$.
		By definition of $\mathcal{X}_{\xi}$ and the uniqueness of the solution of $\dot{x}=f(x)$, for all $t\leq 0$, there exists $x\in \mathcal{X}_{\xi}$ such that $\tau(x) =t$. Thus, for all $t \geq  0$, we have
		\begin{align}\label{eqn:immersion_backward}
			F(\varphi(-t,\xi)) = \psi(-t, F(\xi)). 
		\end{align}
		Thus, the solution $\psi(-t,F(\xi))$ of $\dot{z} =-g(z)$ is well defined and equal to $F(\varphi(-t,\xi))$ for all $t \geq 0$.  
		\hfill $\Box$
	\end{pf}
	By Proposition \ref{prop:time_reverse}, if we can show that a one-to-one linear immersion does not exist for the time-reversed system $\dot{x}= -f(x)$ over a forward-invariant subdomain of $\mathcal{X}$,  then there is no one-to-one linear immersion for the original system $\dot{x} =f(x)$ over $\mathcal{X}$.  The following example demonstrates how to extend our results to systems with diverging trajectories by combining the above two propositions with Theorem \ref{main_theorem}.
	
	\begin{figure}[]
		\centering
		\begin{subfigure}{0.45\linewidth}
			\centering
			\includegraphics[width=1\textwidth]{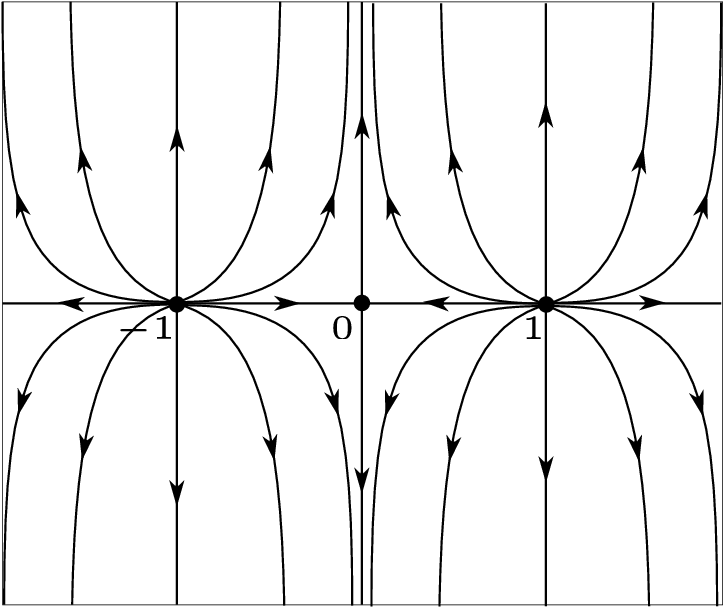}
			\caption{$\dot{x} = f(x)$}
			\label{fig:unbound_example_forward}
		\end{subfigure}
		\begin{subfigure}{0.45\linewidth}
			\centering
			\includegraphics[width=1\textwidth]{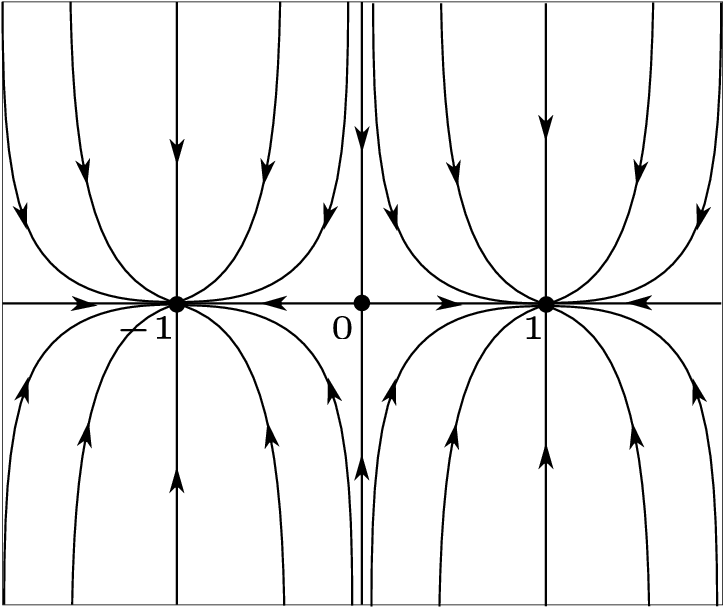}
			\caption{$\dot{x} = -f(x)$}
			\label{fig:unbound_example_backward}
		\end{subfigure}%
		\caption{Phase portraits of the system in Example \ref{ex_6} and its time-reversed counterpart. }
		\label{fig:unbound_example}
	\end{figure}
	
	\begin{example}\label{ex_6}
		Consider a system in $\mathbb{R}^{2}$ with the phase portrait shown in Fig. \ref{fig:unbound_example_forward}. The system has three $\omega$-limit sets $\{-1, 0, 1\}$, where $-1$ and $1$ are unstable equilibria and $0$ is a saddle point. We want to show that there is no one-to-one linear immersion $F$ of this system over $\mathcal{X} = \mathbb{R}^{2}$. 
		To achieve this goal, we cannot directly apply Corollary \ref{cor:linear_system} since any trajectory starting outside $[-1,1]\times \{0\}$ is not precompact. However, by applying Theorem \ref{main_theorem} to the forward-invariant subdomain $\mathcal{X}' =  [-1,0]\times \{0\} $ (or $[0,1]\times \{0\}$) of $\mathcal{X}$, we know that there is no one-to-one linear immersion over $\mathcal{X}'$. Then, by Proposition \ref{prop:extend_X},  no one-to-one linear immersion exists over $\mathcal{X}$. 
		
		To make this example more challenging, consider $\mathcal{X} = \mathbb{R}^{2}\backslash \left( (-1,-0.5)\cup(0.5,1) \right)\times \{0\}$. Note that our previous argument does not work anymore since any forward invariant $\mathcal{X}' \subseteq \mathcal{X}$ contains at least one trajectory that is not precompact. Now consider the time-reversed system, shown by the phase portrait in Fig. \ref{fig:unbound_example_backward}. Since $\mathcal{X}$ is not forward invariant for the time-reversed system, we take the forward invariant subset $\mathcal{X}' = \mathbb{R}^{2}\backslash \left( (-1,0)\cup(0,1) \right)\times \{0\}$ of $\mathcal{X}$. Since all trajectories of the time-reversed systems in $\mathcal{X}'$ are precompact, Corollary \ref{cor:linear_system} implies that there is no one-to-one linear immersion for the time-reversed system over $\mathcal{X}'$. Then, by Proposition \ref{prop:time_reverse}, we know there is no one-to-one linear immersion for the original system over $\mathcal{X}$.
	\end{example}
	
	\subsection{A Direct Extension of Theorem \ref{main_theorem}}\label{sec:direct}
	While Propositions \ref{prop:extend_X} and \ref{prop:time_reverse} allow us to apply Theorem \ref{main_theorem} to systems with diverging trajectories, identifying the appropriate subdomain $\mathcal{X}'$ for more complex examples can be challenging. In this section, we present a direct extension to Theorem  \ref{main_theorem}, where we replace the original condition (T2) with a weaker condition. Before delving into this extension, we introduce several key definitions.
	
	For a system $\dot{x} = f(x)$ defined over $\mathcal{X}$, let $\mathcal{X}'$ be the set of states $\xi\in \mathcal{X}$ such that $\varphi(-t,\xi)$ is defined and contained in $\mathcal{X}$ for all $t\geq 0$ (Namely $\mathcal{X}'$ is the largest backward invariant subset of $\mathcal{X}$ with respect to the system). Given an initial state $\xi\in \mathcal{X}$, the trajectory $\varphi(t, \xi)$ from $\xi$ is called \emph{backward precompact} in $\mathcal{X}$ if $\xi\in \mathcal{X}'$ and the trajectory from $\xi$ is precompact in $\mathcal{X}$ with respect to the time-reversed system $\dot{x} =-f(x)$.

	For each initial state $\xi\in \mathcal{X}$, the $\omega$-limit set of $\xi$ in $\mathcal{X}$ with respect to the time-reversed system $\dot{x}=-f(x)$ is denoted by $\omega^-(\xi)$. If $\xi\not\in \mathcal{X}'$,  we define $\omega^-(\xi)=\emptyset$. The set $\omega^-(\xi)$ is known as an \emph{$\alpha$-limit set} of the original system $\dot{x}=f(x)$ \cite{hirsch2012differential}. For any subset $\Gamma$ of $\mathcal{X}$, its \emph{domain of attraction} with respect to the time-reversed system is denoted by $D^-(\Gamma)$.  We denote the set of all  nonempty $\alpha$-limit sets in $\mathcal{X}$ by  $\mathcal{W}^-$.  Similar to $\mathcal{W}^+$,  the set $\mathcal{W}^-$ contains all the equilibria and closed orbits in $\mathcal{X}$, and thus typically the intersection of $\mathcal{W}^+$ and $\mathcal{W}^-$ is not empty. 
	Now we are ready to present the extension of Theorem \ref{main_theorem}: 
	\begin{theorem}\label{main_theorem_2}
		Suppose that:
		\begin{enumerate}
			\item[(T1')]
			$\dot x=f(x)$ on $\mathcal{X}$ is immersed in a system $\dot{z} = g(z)$ on $\mathcal{Z}$ by a continuous mapping $F$, where both $\dot{z} = g(z)$ and its time-reversed counterpart $\dot{z} = -g(z)$ have closed basins;
			\item[(T2')]
			every trajectory of $\dot{x}=f(x)$ on $\mathcal{X}$ is either precompact or backward precompact in $\mathcal{X}$;
			\item[(T3')]
			the set ${\mathcal{W}^+ \cup \mathcal{W}^-}$ is countable.
		\end{enumerate}
		Then, the set $\{F(\Omega) \mid  \Omega\in \mathcal{W}^+ \cup \mathcal{W}^-\} $ has exactly one maximal element, that is, there exists $\Omega\in \mathcal{W}^+ \cup \mathcal{W}^-$ such that $F(\Omega)\supseteq F(\Omega')$ for all $\Omega'$ in $\mathcal{W}^+\cup \mathcal{W}^-$.
		
		If in addition to (T1')-(T3'), we also have:
		\begin{enumerate}
			\item[(T4')] For every $\Omega^+\in \mathcal{W}^+$ and $\Omega^-\in \mathcal{W}^-$, there exist at least one precompact trajectory in $D^+(\Omega^+)$ and one backward precompact trajectory in $D^-(\Omega^-)$,
		\end{enumerate}
		then the set $\{F(\Omega) \mid  \Omega\in \mathcal{W}^+ \cup \mathcal{W}^-\} $ has exactly one element, that is $F(\Omega_1) = F(\Omega_2)$ for any $\Omega_1$ and $\Omega_2$ in $\mathcal{W}^+ \cup \mathcal{W}^-$. 
	\end{theorem}
	The proof of Theorem \ref{main_theorem_2} is similar to that of Theorem \ref{main_theorem}, and can be found in Appendix \ref{app:extension}. 
	Under conditions (T1')-(T3'), Theorem \ref{main_theorem_2} says that any continuous immersion $F$ cannot fully distinguish different limit sets in  $\mathcal{X}$. 
	Compared with Theorem \ref{main_theorem}, (T2') is relaxed to allow diverging trajectories, as long as these trajectories converge to some limit sets (such as an unstable equilibrium) backward in time. At the same time, Theorem \ref{main_theorem_2} requires that the time-reversed system $\dot{z}=-g(z)$ also has closed basins, which is trivially satisfied by any linear system. 
	
	\begin{example}\label{ex_7}
		Consider the same system in Example \ref{ex_6} and $\mathcal{X} = \mathbb{R}^{2}\backslash \left( (-1,-0.5)\cup(0.5,1) \right)\times \{0\}$. Since (T2')-(T4') in Theorem \ref{main_theorem_2} are satisfied, any linear immersion over $\mathcal{X}$ collapses the three equilibria into one.  Therefore, no one-to-one linear immersions exist over this specific $\mathcal{X}$. Compared with the indirect extensions of Theorem \ref{main_theorem}, Theorem \ref{main_theorem_2} is directly applied to show the non-existence of linear immersions, without constructing the subdomain $\mathcal{X}'$ as in Example \ref{ex_6}.
	\end{example}
	
	\section{Conclusion} \label{sec:conclusion} 
	In this work, we first show that linear immersions collapse different $\omega$-limit sets into one under the condition that (i) all trajectories in $\mathcal{X}$ are precompact and (ii) there are at most countably many $\omega$-limit sets. Then we bridge our theoretical findings on exact linear immersions with approximate linear immersions learned from data. We show that as the size of the data set increases and the sampling interval decreases, the learned linear immersion converges to functions incapable of distinguishing different $\omega$-limit sets. To extend the applicability of our results beyond the constraints of precompact trajectories, we have also presented several extensions to the main theorem. These extensions broaden the scope of our results, allowing us to address systems with diverging trajectories. For future work, we are interested in the effect of multiple $\omega$-limit sets on   the approximation error of Koopman-based projected models \cite{haseli2023invariance}, which is well-defined even if a linear immersion does not exist.
	\begin{ack}                               % Place acknowledgements
		We thank Matthew Kvalheim for helpful discussions about the relation of our results with those in \cite{kvalheim2023linearizability}.
	\end{ack}
	\bibliographystyle{abbrv}
	\bibliography{ref}
	\appendix
	
	\section{Proof of the Main Theorem} \label{sec:proof} 
	To prove Theorem \ref{main_theorem}, we first need to introduce two lemmas. Recall that by Remark \ref{continuity}, an immersion is assumed to be continuous unless otherwise specified. The first lemma reveals a relation between $\omega$-limit sets of the original system and the immersed system.
	\begin{lemma}\label{map_limit_sets} Let $F$ be an immersion that maps $\dot{x}=f(x)$ on $\mathcal{X}$ to $\dot{z}=g(z)$ on $\mathcal{Z}$. For any $\xi\in \mathcal{X}$, if $\omega^+(\xi)$ is nonempty, then $\omega^+(F(\xi))$ exists and contains $F(\omega^+(\xi))$. 
		Furthermore, if the trajectory starting from $\xi$ is {precompact in $\mathcal{X}$},  $F(\omega ^+(\xi )) = \omega ^+(F(\xi ))$.
	\end{lemma}
	
	{\bf Proof.}
	We first prove that $F(\omega ^+(\xi )) \subseteq  \omega ^+(F(\xi ))$.
	Indeed, suppose that $p\in \omega ^+(\xi )$, and pick a sequence of times $t_i\rightarrow \infty $ so
	that $\varphi(t_i,\xi )\rightarrow p$ as $t_i\rightarrow \infty $. 
	Therefore $\psi (t_i,F(\xi )) = F(\varphi(t_i,\xi ))\rightarrow q:=F(p)$, showing that
	$q\in \omega ^+(F(\xi ))$.
	
	Conversely, suppose that $q'\in \omega ^+(F(\xi ))$, and pick a sequence of times
	$t_i\rightarrow \infty $ so that $F(\varphi(t_i,\xi )) = \psi (t_i,F(\xi )) \rightarrow q'\in \mathcal{Z}$ as $t_i\rightarrow \infty $. 
	{Since the trajectory $\varphi(t,\xi )$ is precompact in $\mathcal{X}$}, there is a subsequence of the $t_i$'s, which is again denoted by $t_i$
	without loss of generality, so that $\varphi(t_i,\xi )\rightarrow p\in \mathcal{X}$ and thus
	$\psi (t_i,F(\xi ))=F(\varphi(t_i,\xi ))  \rightarrow q:=F(p)$.
	Since we picked a subsequence, also $\psi (t_i,F(\xi )) = F(\varphi(t_i,\xi ))\rightarrow q'$.
	We conclude that $q'=q\in F(\omega ^+(\xi ))$, showing that
	$\omega ^+(F(\xi ))\subseteq F(\omega ^+(\xi ))$.
	We conclude that $F(\omega ^+(\xi )) = \omega ^+(F(\xi ))$.
	\hfill $\Box$
	
	\begin{remark}\label{map_all_lmit_sets}
		{Under condition (T2) in Theorem \ref{main_theorem}}, for any $\Omega \in \mathcal{W}^+$, the image $\hat\Omega :=F(\Omega )$ is an $\omega$-limit set for the
		system $\dot z=g(z)$.
		Indeed, by definition there is some $\xi \in \mathcal{X}$ such that $\omega ^+(\xi )=\Omega $. Thus, from
		Lemma~\ref{map_limit_sets} we have that
		$\hat\Omega  = F(\Omega ) = F(\omega ^+(\xi )) = \omega ^+(F(\xi ))$.
	\end{remark}
	
	Next, observe that, in general, $F(D^+(\Omega ))\not= D^+(F(\Omega ))$, since the latter set could be
	larger.  Examples are easy to construct by taking $\mathcal{X}$ to be a
	forward-invariant subset of $\mathcal{Z}$ and $F$ the identity.
	For example, consider $\dot x=-x$ on
	$\mathcal{X}=(-1,1)$ and the same system $\dot z=-z$ on $\mathcal{Z}=\mathbb{R}$.
	Here $\Omega =\{0\}$ is the only $\omega$-limit set, and $F(D^+(\Omega )) = D^+(\Omega )=(-1,1)$ but
	$D^+(F(\Omega ))=\mathbb{R}$.
	However, the following weaker statement is true.
	\begin{lemma}\label{inverse_image}
		Suppose that $F$ is an immersion. For any $\xi \in \mathcal{X}$, if $\varphi(t,\xi)$ is precompact in $\mathcal{X}$, then $F(D^+(\omega^+(\xi))) \subseteq  D^+(F(\omega^{+}(\xi)))$.
	\end{lemma}
	
	{\bf Proof.} Since $\varphi(t,\xi)$ is precompact in $\mathcal{X}$, by Lemmas \ref{lem:precompact} and  \ref{map_limit_sets}, $\Omega:=\omega^+(\xi)$ is nonempty and $F(\Omega) = \omega^+(F(\xi))$. Let $x$ be a point in $D^+(\Omega)$. Then, there exists a sequence $t_n\geq 0$ such that $t_n \rightarrow + \infty$ and $\varphi(t_n, x)\rightarrow x'$ for some $x'\in \Omega$.  By \eqref{eqn:immersion} and the continuity of $F$, $\lim_{n\rightarrow +\infty} \psi(t_n, F(x)) = \lim_{n\rightarrow +\infty}  F(\varphi(t_n, x)) = F(x')\in F(\Omega) = \omega^+(F(\xi))$. Hence, $F(x)\in D^+( \omega^+(F(\xi)))$ and thereby $F(D^+(\omega^+(\xi)))\subseteq D^+(F(\omega^+(\xi)))$. \hfill $\Box$ 
	
	{\bf Proof of Theorem \ref{main_theorem}:}
	Since by (T2) every trajectory is precompact {in $\mathcal{X}$} and by (T3) there are at most countably many $\omega$-limit sets in $\mathcal{X}$, we have that $\mathcal{X}=\bigcup _{i\in I} D^+(\Omega _i)$, for
	a countable set $I$.  Thus, $F(\mathcal{X}) = \bigcup_{i\in I}F(D^{+}(\Omega_{i}))$.
	By (T2) and Lemmas \ref{map_limit_sets} and \ref{inverse_image}, $F(\Omega_i)$ is an $\omega$-limit set for all $i\in I$ and 
	\begin{align}\label{eqn:union_F}
		F(\mathcal{X}) = \bigcup_{i\in I} \left( D^{+}(F(\Omega_{i}))\cap F(\mathcal{X}) \right).
	\end{align}
	According (T1), $D^{+}(F(\Omega_{i}))$ is closed  in  $\mathcal{Z}$ and thus $D^{+}(F(\Omega_{i}))\cap F(\mathcal{X})$ is closed in the subspace topology induced on $F(\mathcal{X})$ for all $i\in I$. Moreover, $D^{+}(F(\Omega_{i}))\cap F(\mathcal{X})$ for all $i\in I$ is nonempty since points in $F(\Omega_i)$ are limit points of $D^{+}(F(\Omega_{i}))$ and thus contained in the closed set $D^{+}(F(\Omega_{i}))\cap F(\mathcal{X})$. 
	
	Finally, since the domains of attraction of two different $\omega$-limit sets must be disjoint, for any indices $i$ and $i'\in I$,  we have either 
		(a) the two $\omega$-limit sets $F(\Omega_{i})$ and $F(\Omega_{i'})$ of the immersed system are equal, which implies $D^{+}(F(\Omega_{i}))=D^{+}(F(\Omega_{i'}))$; or (b) $D^{+}(F(\Omega_{i}))$ and $D^{+}(F(\Omega_{i'}))$ are disjoint. 
		
		Suppose there exist $i$ and $i'$ such that case (b) holds, that is, $D^{+}(F(\Omega_{i}))$ and $D^{+}(F(\Omega_{i'}))$ are disjoint. It follows from \eqref{eqn:union_F} that $F(\mathcal{X})$ is a disjoint union of a countable collection of closed sets. Since $\mathcal{X}$ is path-connected and $F$ is continuous, $F(\mathcal{X})$ is path-connected as well. 
		Thus, by the main theorem of \cite{sierpinski1918theoreme}, only one of the sets in the collection $\{D^{+}(F(\Omega_{i})) \cap F(\mathcal{X})\}_{i\in I}$ can be nonempty. This contradicts our earlier proof that $D^{+}(F(\Omega_{i})) \cap F(\mathcal{X})$ is nonempty for all $i \in I$. Thus by contradiction, $F(\Omega_{i})$ must be the same for all $i\in I$. \hfill $\Box$
	~\\

	\begin{remark}\label{sierpinski}
		Sierpi\'nski's Theorem states that if a continuum $\mathcal{X}$ has a
		countable cover $\{X_i\}_{i=1}^{\infty }$ by pairwise disjoint closed subsets, then at
		most one of the sets $X_i$ is non-empty. A continuum is a compact connected
		Hausdorff space, but we do not assume that $\mathcal{X}$ is compact. However, the
		theorem is still true if $\mathcal{X}$ is not compact. Indeed, suppose that two of the sets $X_i$ would be
		nonempty, and pick two points $p,q$, one in each set. Consider a (continuous)
		path $\gamma :[0,1]\rightarrow \mathcal{X}$ that joins these two points, and let $\Gamma :=\gamma ([0,1])$.
		Now the sets $\{X_i\cap \Gamma \}_{i=1}^{\infty }$ form a disjoint cover of the continuum
		$\Gamma $, but two of these sets are nonempty, a contradiction.
	\end{remark}
	\section{Proof of Theorem \ref{main_theorem_2}}\label{app:extension}
	
	Recall that the definitions of backward precompact trajectories, $\alpha$-limit sets, and their domains of attraction are provided in Section \ref{sec:direct}. 
	We first show a property of systems with closed basins.
	\begin{lemma}\label{lem:omega_alpha_eq}
		Suppose that both the original system $\dot{x}=f(x)$ and the time-reversed system $\dot{x}=-f(x)$ have closed basins.
		Then, for any $\omega$-limit set $\Omega \subseteq \mathcal{X}$ and $\alpha$-limit set $\Gamma \subseteq \mathcal{X}$ of $\dot{x}=f(x)$, $D^+(\Omega)\cap D^-(\Gamma) \not=\emptyset$ implies $\Omega=\Gamma$.
	\end{lemma}
	\begin{pf}
		Let $x_0 \in D^+(\Omega)\cap D^-(\Gamma) $. {That is, $\omega^+(x_0)=\Omega$ and $\omega^-(x_0)=\Gamma$.} Denote the trajectory through $x_0$ by $X_0=\{\psi(t,x_0) \mid t\in \mathbb{R} \}$. Then, by the definition of limit sets, {$\omega^+(x)=\Omega$ and $\omega^-(x)=\Gamma$ imply that (i) $X_0$ is contained by $  D^+(\Omega)\cap D^-(\Gamma)$ and (ii) the closure $\text{cl}(X_0)$ contains $\Omega\cup\Gamma$.} Since $D^+(\Omega) $ and $ D^-(\Gamma) $ are closed, we have $$\Omega \cup \Gamma\subseteq \text{cl}(X_0)\subseteq D^+(\Omega)\cap D^-(\Gamma).$$
		
		Thus, we have $\Omega \subseteq D^-(\Gamma)$ and $\Gamma \subseteq D^+(\Omega)$.
		Next, it can be shown that any limit set is closed and invariant in $\mathcal{X}$ \cite{alligood2000chaos}.
		Since $\Omega$ is invariant and $\Omega \subseteq D^-(\Gamma)$, we have $\Gamma \subseteq \text{cl}(\Omega) = \Omega$. Similarly, we have $\Omega \subseteq \text{cl}(\Gamma)=\Gamma$.  Thus, $\Omega = \Gamma$. \hfill $\Box$
	\end{pf}

	Next, we extend Lemmas \ref{map_limit_sets} and \ref{inverse_image} for $\alpha$-limit sets.
	
	\begin{lemma}\label{map_alpha_limit_sets} Let $F$ be an immersion that maps $\dot{x}=f(x)$ on $\mathcal{X}$ to $\dot{z}=g(z)$ on $\mathcal{Z}$. 
		For any $\xi \in \mathcal{X}$, if $\omega^-(\xi)$ exists, then $\omega^-(F(\xi))$ exists and contains $F(\omega^-(\xi))$. Furthermore, if the trajectory starting at $\xi$ is backward precompact in $\mathcal{X}$,  ${F(\omega^-(\xi )) = \omega^-(F(\xi ))}$.
	\end{lemma}
	\begin{pf} Let $\mathcal{X}'$ be the maximal forward invariant subset of $\mathcal{X}$ with respect to the time-reversed system.   By Proposition \ref{prop:time_reverse}, $F$ is an immersion that maps $\dot{x}=-f(x)$ on $\mathcal{X}'$ to $\dot{z}=-g(z)$ on $F(\mathcal{X}')$. Then, the proof is completed by applying Lemma \ref{map_limit_sets} to the time-reversed system $\dot{x} = -f(x)$ on $\mathcal{X}'$ and the immersion $F$. \hfill $\Box$
	\end{pf}
	
	\begin{lemma}\label{inverse_image_2}
		Let $F$ be an immersion that maps $\dot{x}=f(x)$ on $\mathcal{X}$ to  a system $\dot{z}=g(z)$ on $\mathcal{Z}$.
		For any $\xi\in \mathcal{X}$, if $\varphi(t,\xi)$ is precompact in $\mathcal{X}$, then 
		$$F(D^+(\omega^+(\xi))) \subseteq  D^+(F(\omega^+(\xi))),$$ 
		and if $\varphi(t,\xi)$ is backward precompact in $\mathcal{X}$, then
		$$F(D^-(\omega^-(\xi))) \subseteq  D^-(F(\omega^-(\xi) )).$$\end{lemma}
	
	\begin{pf}  
		{The inclusion relation when $\varphi(t,\xi)$ is precompact in $\mathcal{X}$ is directly implied by Lemma \ref{inverse_image} .
			The inclusion relation when $\varphi(t,\xi)$ is backward precompact in $\mathcal{X}$ can be shown via similar arguments in the proof of Lemma \ref{inverse_image} with Lemma \ref{lem:precompact}, Lemma \ref{map_alpha_limit_sets}, and \eqref{eqn:immersion_backward}. }\hfill $\Box$
	\end{pf}

	The following lemma is a key component in the proof of Theorem~\ref{main_theorem_2}.

		\begin{lemma} \label{lem:F_equal} 
			Let $\widehat{\mathcal{W}}^+$ be the set of $\omega$-limit sets $\Omega$ in $\mathcal{W}^+$ whose domain of attraction $D^+(\Omega)$ contains at least one trajectory precompact in $\mathcal{X}$, and let $\widehat{\mathcal{W}}^-$ be the set of $\alpha$-limit sets $\Gamma$ in $\mathcal{W}^-$ whose domain of attraction $D^-(\Gamma)$ contains at least one trajectory backward precompact in $\mathcal{X}$. Then, 
			under (T1')-(T3'), the images $F(\Omega)$ and $F(\Omega')$ are the same for any $\Omega$ and  $\Omega'$ in $\widehat{\mathcal{W}}^{+}\cup\widehat{\mathcal{W}}^{-}$.
		\end{lemma}

		\begin{pf} For any limit set $\Omega\in \widehat{\mathcal{W}}^+ \cup \widehat{\mathcal{W}}^-$, let us define
			\begin{align} \label{eqn:D} 
				D( \Omega) &= \begin{cases}
					D^{+}(\Omega) & \Omega \in \widehat{\mathcal{W}}^+ \backslash \widehat{\mathcal{W}}^-\\
					D^{-}(\Omega) & \Omega \in \widehat{\mathcal{W}}^- \backslash \widehat{\mathcal{W}}^+\\
					D^{+}(\Omega)\cup D^{-}(\Omega) & \Omega \in \widehat{\mathcal{W}}^- \cap \widehat{\mathcal{W}}^+
				\end{cases},\\
				D( F(\Omega)) &=  \begin{cases}
					D^{+}(F(\Omega)) & \Omega \in \widehat{\mathcal{W}}^+ \backslash \widehat{\mathcal{W}}^-\\
					D^{-}(F(\Omega)) & \Omega \in \widehat{\mathcal{W}}^- \backslash \widehat{\mathcal{W}}^+\\
					D^{+}(F(\Omega))\cup D^{-}(F(\Omega)) & \Omega \in \widehat{\mathcal{W}}^- \cap \widehat{\mathcal{W}}^+
				\end{cases}.\nonumber
			\end{align}
			Now, we show that for any $\Omega,\ \Omega'\in \widehat{\mathcal{W}}^+\cup\widehat{\mathcal{W}}^- $ such that $F(\Omega) \not= F(\Omega')$, $D(F(\Omega))$ and $D(F(\Omega'))$ must be disjoint.
			
			Assume that $F(\Omega)\not=F(\Omega')$. Immediately, we have 
			\begin{align}
				\begin{split} \label{eqn:DoA_disjoint} 
					D^{+}(F(\Omega))\cap D^{+}(F(\Omega')) = \emptyset,\\
					D^{-}(F(\Omega))\cap D^{-}(F(\Omega')) = \emptyset,\\
					D^{+}(F(\Omega))\cap D^{-}(F(\Omega')) = \emptyset,\\
					D^{-}(F(\Omega))\cap D^{+}(F(\Omega')) = \emptyset,
				\end{split}
			\end{align}
			where the first two equations hold by the definition of domain of attraction, and the last two equations hold due to (T1') and Lemma \ref{lem:omega_alpha_eq}. 
			By \eqref{eqn:D} and \eqref{eqn:DoA_disjoint}, if $F(\Omega)\not= F(\Omega')$, $D(F(\Omega))$ and $D(F(\Omega'))$ must be disjoint.
			
			Now we are ready to show that $F(\Omega) = F(\Omega')$ for all $\Omega,\ \Omega'\in \widehat{\mathcal{W}}^+\cup\widehat{\mathcal{W}}^-$ under conditions (T1')-(T3').
			
			Since by (T2'), for any $\xi\in \mathcal{X}$, the corresponding trajectory is either precompact or backward precompact in $\mathcal{X}$, by Lemma \ref{lem:precompact}, any $\xi\in \mathcal{X}$ must belong to $D(\Omega)$ for some $\Omega\in \widehat{\mathcal{W}}^+\cup \widehat{\mathcal{W}}^-$. Thus, the set $\mathcal{X} = \cup_{i\in I} D(\Omega_{i})$ for a countable set $I$, where $\{\Omega_{i}\}_{i\in I} = \widehat{\mathcal{W}}^+ \cup \widehat{\mathcal{W}}^- $. As a result, $F(\mathcal{X}) = \cup_{i\in I} F(D(\Omega_{i}))$. By Lemma \ref{inverse_image_2}, $F(D(\Omega_{i})) $ is contained by $ D(F(\Omega_{i}))$. Thus, 
			\begin{align} \label{eqn:union_F_2}
				F(\mathcal{X}) = \bigcup_{i\in I} (D(F(\Omega_{i})) \cap F(\mathcal{X})).
			\end{align}
			By (T1'), $D(F(\Omega_{i})) \cap F(\mathcal{X})$ is closed in the subspace topology induced by $F(\mathcal{X})$ for all $i\in I$. Moreover, the set $ D(F(\Omega_{i})) \cap F(\mathcal{X}) $ is nonempty  for all $i\in I$ since points in $F(\Omega_i)$ are limit points of $D(F(\Omega_{i}))$ and thus contained in the closed set $D(F(\Omega_{i}))$ (as both $\dot{z}=g(z)$  and $\dot{z}=-g(z)$ have closed basins).
			
			Now suppose that $F(\Omega_{i})$ and $F(\Omega_{j})$ are not equal for some $i$ and $j$ in $I$. Then, $D(F(\Omega_{i}))$ and $D(F(\Omega_{j}))$ are nonempty and disjoint, and thus \eqref{eqn:union_F_2} implies that $F(\mathcal{X})$ is a disjoint union of a countable collection of closed sets.  Since $\mathcal{X}$ is path-connected and $F$ is continuous, $F(\mathcal{X})$ is path-connected. Thus, by the main theorem of \cite{sierpinski1918theoreme}, only one of the sets in the collection $\{D(F(\Omega_{k})) \cap F(\mathcal{X})\}_{k\in I}$ can be nonempty. This contradicts our earlier proof that $ D(F(\Omega_i)) \cap F(\mathcal{X})$ and $ D(F(\Omega_j)) \cap F(\mathcal{X})$ are disjoint and nonempty. Thus, by contradiction, the sets $F(\Omega_{k})$ must be the same for all $k\in I$. That is, $F(\Omega)$ is the same set for all $\Omega\in \widehat{\mathcal{W}}^+ \cup \widehat{\mathcal{W}}^-$. \hfill $\Box$
		\end{pf}

		\textbf{Proof of Theorem \ref{main_theorem_2}:} Recall under (T1')-(T3'), we want to show that there exists $\Omega_{\max} \in \mathcal{W}^{+}\cup \mathcal{W}^{-}$ such that $F(\Omega_{\max}) \supseteq F(\Omega)$ for any $\Omega \in \mathcal{W}^{+}\cup \mathcal{W}^{-}$. Note that (T2') and Lemmas \ref{map_limit_sets} and \ref{map_alpha_limit_sets} imply that $\widehat{\mathcal{W}}^{+}\cup \widehat{\mathcal{W}}^{-}$ is nonempty.  Let $\Omega_{\max}$ be any element in the nonempty set $\widehat{W}^{+}\cup \widehat{W}^{-}$.  For any $\Omega\in \mathcal{W}^{+} \cup \mathcal{W}^{-}$, we show $F(\Omega) \subseteq F(\Omega_{\max})$ by considering three cases: 
		
		\underline{Case 1:} $\Omega\in \widehat{\mathcal{W}}^{+}\cup \widehat{\mathcal{W}}^{-}$. By Lemma \ref{lem:F_equal}, $F(\Omega) = F(\Omega_{\max})$.  
		
		\underline{Case 2:} $\Omega \in \mathcal{W}^{+}\backslash (\widehat{\mathcal{W}}^{+}\cup \widehat{\mathcal{W}}^{-})$. Pick any $\xi\in D^{+}(\Omega)$. Since $\Omega \not \in \widehat{\mathcal{W}}^{+}$, the solution of $\xi$ cannot be precompact in $\mathcal{X}$, and thus must be backward precompact by (T2'). Thus,  by Lemma \ref{lem:precompact}, $\omega^{-}(\xi)$ exists and belongs to $\widehat{W}^{-}$. By Lemmas \ref{map_alpha_limit_sets} and \ref{lem:F_equal}, $F(\omega^{-}(\xi)) = \omega^{-}(F(\xi)) = F(\Omega_{\max})$. By Lemma \ref{map_limit_sets}, $\omega^{+}(F(\xi))$ is nonempty and contains $F(\Omega)$. Finally, by Lemma \ref{lem:omega_alpha_eq} and (T1'), we have $\omega^{+}(F(\xi)) = \omega^{-}(F(\xi))$. Therefore, we have
		\begin{align}
			F(\Omega) \subseteq \omega^{+}(F(\xi))= \omega^{-}(F(\xi)) = F(\Omega_{\max}).
		\end{align}
		\underline{Case 3:} $\Omega\in \mathcal{W}^{-}\backslash (\widehat{\mathcal{W}}^{+}\cup \widehat{\mathcal{W}}^{-}) $. By arguments similar to Case 2, it can be shown that $F(\Omega) \subseteq F(\Omega_{\max})$. 
		
		So far, we have shown that under conditions (T1')-(T3'), $F(\Omega) \subseteq F(\Omega_{\max})$ for any $\Omega \in \mathcal{W}^{+}\cup \mathcal{W}^{-}$. Finally, note that (T4') implies $\mathcal{W}^{+} = \widehat{\mathcal{W}}^{+}$ and $\mathcal{W}^{-} = \widehat{\mathcal{W}}^{-}$, and thus for any $\Omega \in \mathcal{W}^{+}\cup \mathcal{W}^{-}$, $F(\Omega) = F(\Omega_{\max})$ by Lemma \ref{lem:F_equal}. \hfill $\Box$
	
\end{document}